\documentclass[11pt,a4paper]{emulateapj}
\usepackage{graphicx}        
\usepackage{amsmath}
\usepackage{color}           
\usepackage{url}             

\usepackage{epsfig}
\usepackage{amsmath}

\newcommand{\beq}{\begin{equation}}
\newcommand{\eeq}{\end{equation}}
\newcommand{\beqa}{\begin{eqnarray}}
\newcommand{\eeqa}{\end{eqnarray}}

\begin{document}



\title{Numerical simulations of multi-shell plasma twisters 
in the solar atmosphere
}

\author{K.~Murawski$^{1}$,
        A.K.~Srivastava$^{2}$,
        Z.E.~Musielak$^{3,4}$,
	B.N.~Dwivedi$^{2}$
       }

\shortauthors{K.~Murawski et al.}
\shorttitle{Multi-shell solar twisters}

\altaffiltext{1}{Group of Astrophysics, Institute of Physics, UMCS, ul. Radziszewskiego 10, 20-031 Lublin, Poland}
\altaffiltext{2}{Department of Physics, Indian Institute of Technology (Banaras Hindu University), Varanasi-221005, India}
\altaffiltext{3}{Department of Physics, University of Texas at Arlington, Arlington, TX 76019, USA}
\altaffiltext{4}{Kiepenheuer-Institut f\"ur Sonnenphysik, Sch\"oneckstr. 6, 79104 Freiburg, Germany}

\begin{abstract}
We perform numerical simulations of impulsively generated Alfv\'en waves in 
an isolated photospheric flux tube, and 
explore 
the propagation of these waves 
along such magnetic structure that extends from the photosphere, where these 
waves are 
triggered, 
to the solar corona, and analyze resulting magnetic shells.  
Our model of the solar atmosphere is constructed by adopting the temperature 
distribution based on the semi-empirical model and specifying the curved 
magnetic field lines that constitute the magnetic flux tube which is rooted in the solar photosphere. 
The evolution of the solar atmosphere is described 
by 3D, ideal magnetohydrodynamic equations that are numerically solved by the FLASH code. 
Our numerical simulations reveal, based on the physical properties of the multi-shell magnetic twisters 
and the amount of energy and 
momentum associated with 
them, that these multi-shell magnetic twisters may be responsible 
for the observed heating of the 
lower 
solar corona and for the formation of 
solar wind. 
Moreover, it is likely that the existence of these 
twisters can be verified by high-resolution observations. 
\end{abstract}


%
\section{Introduction}
%
The existence of hot solar corona ($\ge 10^6$ K) is well-established by observations of 
solar ultraviolet and X-ray emissions originating in magnetically confined coronal plasma, 
and the acceleration of fast solar wind 
that 
commence from 
magnetically open coronal regions (Peter and Dwivedi 2014). 
The identification of physical processes responsible for the coronal heating and solar wind acceleration 
has been a very challenging and still remains unsolved problem (Priest et al. 1998, Cranmer 2002, Peter and Dwivedi 2014). 
Recent high-resolution solar observations seem to clearly imply that the energy and mass transport required to heat 
the corona and accelerate the solar wind must 
come from 
the chromosphere through highly localized wave and 
dynamical plasma processes at small spatio-temporal scales, such as spicules, network jets, swirls, 
and twisting motions of magnetized plasma (De Pontieu et al. 2004, Shibata et al. 2007, De Pontieu et al. 2014, McIntosh et al. 2011). 

Solar chromospheric swirls have been detected by recent observations (Wedemeyer-B\"ohm et al. 2012). 
They are categorized by their various morphological shapes (Tian et al. 2014). 
%
{\bf 
These swirls are suggested to be driven by solar convective motions 
(Wedemeyer-B\"ohm et al. 2012, Steiner et al. 2010, Shelyag et al. 2011). 
}
Magnetically-confined plasma in such swirls couples to the solar upper atmospheric layers, transferring swirls’ rotation into the low corona. 
At higher altitudes, these swirls may appear like large cyclones/tornadoes due to expansion of magnetic flux tubes 
that support them (Shelyag et al. 2011, Su et al. 2012, Zhang and Liu 2011). 
The relationship between the chromospheric swirls and coronal tornadoes is not yet understood 
(Wedemeyer et al. 2013, Shelyag et al. 2013, Wedemeyer and Steiner 2014). 
It is also debated whether the swirls are triggered by solar convective motions, 
or they are associated with torsional Alfv\'en waves at small spatial scales (Shelyag et al. 2013, Wedemeyer and Steiner 2014). 
The observations, however, suggest that these localized motions of magnetized plasma 
{\bf 
might transfer up to 
}
$1.4\times 10^4$  W m$^{-2}$ energy flux to the base of the solar corona for its localized heating and 
wind acceleration (Wedemeyer-B\"ohm et al. 2012). 
 
The main objective of this paper is to describe novel solar phenomena called 
here multi-shell magnetic twisters, and their driving mechanism and related morphology. 
It should be noted that recently observed chromospheric swirls can be an integrated 
appearance of multi-shell magnetic twisters. 
Our finding sheds new light on the morphological evolution of such twisters, 
their exact excitation mechanism and their role in the localized 
heating 
of 
the solar corona, 
and mass transport to nascent solar wind. The novel result is our theoretical prediction 
of the multi-shell magnetic twisters, whose existence can be verified by observations 
with upcoming high-resolution instruments.
 
The paper is organized as follows. Our model of solar magnetic arcades and 
description of our numerical method are introduced in Sects.~2 and~3, respectively. 
Results of our numerical simulations of the Alfv\'en wave 
propagation in magnetic arcades are presented and discussed in Sect.~4. 
Conclusions are given in Sect.~5. 
%
\section{Numerical model of 
multi-shells}\label{sec:atm_model}
%
%
\subsection{MHD equations}\label{sec:equ_model}
%
We consider a gravitationally stratified and magnetically confined plasma in a 
structure that resembles a flux tube, which is described by the following set of 
MHD equations:
%
\beqa
\label{eq:MHD_rho} 
{{\partial \varrho}\over {\partial t}}+\nabla \cdot (\varrho{\bf V})=0\, ,\\
\label{eq:MHD_V}
\varrho{{\partial {\bf V}}\over {\partial t}}+ \varrho\left ({\bf V}\cdot \nabla\right )
{\bf V}= -\nabla p+ \frac{1}{\mu} (\nabla\times{\bf B})\times{\bf B} +\varrho{\bf g}\, , \\
\label{eq:MHD_B}
{{\partial {\bf B}}\over {\partial t}}= \nabla \times ({\bf V}\times {\bf B})\, , \\
\label{eq:MHD_divB}
\nabla\cdot{\bf B} = 0\, , \\
\label{eq:MHD_p}
{\partial p\over \partial t} + {\bf V}\cdot\nabla p = -\gamma p \nabla \cdot {\bf V}\, ,\\
\label{eq:MHD_CLAP}
p = \frac{k_{\rm B}}{m} \varrho T\, ,
\eeqa
%
where ${\varrho}$ is mass density, $p$ gas pressure, ${\bf V}$, ${\bf B}$ and 
${\bf g}=(0,-g, 0)$ represent the plasma velocity, the magnetic field and gravitational 
acceleration, respectively.  In addition, $T$ is temperature, $m$ particle mass,
that was specified by mean molecular weight value of $1.24$ (Oskar Steiner, private communication), 
$k_{\rm B}$ is the Boltzmann's constant, $\gamma=1.4$ the adiabatic index, 
and $\mu$ the magnetic permeability of plasma. 
The value of $g$ is $274$ m s$^{-2}$.

%
\begin{figure}
	\begin{center}
		\includegraphics[scale=0.25, angle=0]{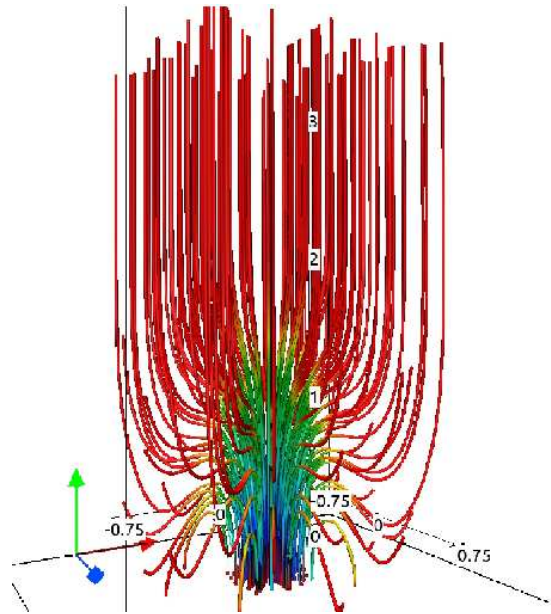}
		\caption{\small Magnetic field lines at $t=0$ s.} 
		\label{fig:mag_e}
	\end{center}
\end{figure}
%



%
%
\begin{figure}
	\begin{center}
		\includegraphics[scale=0.35, angle=0]{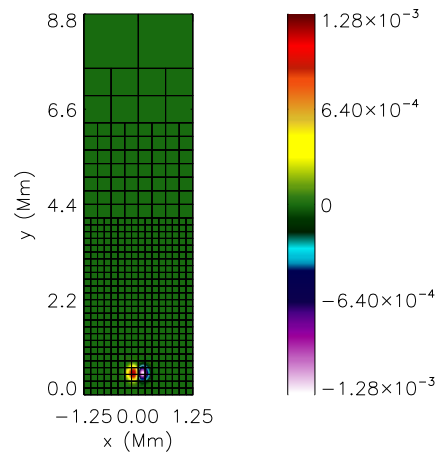}
		\caption{\small Numerical blocks used in the numerical simulations and 
                                the initial pulse, $V_{\rm \theta}(x,y,z=0)$, drawn in the vertical ($x-y$) plane for $z=0$. 
                Note that a part of the whole simulation region is displayed only and $max(\vert V_{\rm \theta}\vert)=2.56$ km s$^{-1}$. 
}
		\label{fig:blk}
	\end{center}
\end{figure}
%

%
\subsection {A model of the static solar atmosphere}\label{sec:equil}
%
We consider a model of the static ($\partial/\partial t = 0$) solar atmosphere 
in which all plasma quantities vary with $x$, $y$, and $z$. 
We assume that the solar atmosphere is in static equilibrium (${\bf V_{\rm e}}={\bf 0}$) 
with the Lorentz force balanced by the gravity force and the gas pressure gradient, 
which means that 
\begin{equation}
\label{eq:force_free}
\frac{1}{\mu}(\nabla\times{\bf B_{\rm e}})\times{\bf B_{\rm e}} + 
\varrho_{\rm e} {\bf g} -\nabla p_{\rm e} = {\bf 0}\ ,
\end{equation}
\noindent
where the subscript $'{\rm e}'$ corresponds to the equilibrium configuration. 
We consider an axisymmetric flux tube of which equilibrium is described by Murawski et al. (2015) 
which is based on the analytical theory developed by Solov'ev (2010). 
This flux tube is initially non-twisted, and 
its magnetic field is represented 
by the azimuthal component of 
the 
magnetic flux function 
($A\,{\bf\hat \theta}$) as
\beq\label{eq:equil_B}
{\bf B_{\rm e}} = \nabla\times (A\,{\bf\hat \theta})\, ,
\eeq
where ${\bf\hat \theta}$ is a unit vector along the azimuthal direction. 
In this case, we have
\beq\label{eq:B_com_A}
B_{\rm er} = -\frac{\partial A}{\partial y}\, ,\hspace{3mm} 
B_{\rm e\theta} = 0\, , \hspace{3mm} 
B_{\rm ey} = \frac{1}{r} \frac{\partial (rA)}{\partial r}\, ,
\eeq
where $r=\sqrt{x^2+z^2}$ is the radius. 
For a flux tube, we can specify $A(r,y)$ as 
\beqa\label{eq:A-Psi-flux tube}
A(r,y) = B_{\rm 0} \exp{(-k_{\rm y}^2 y^2)} \frac{r}{1+k_{\rm r}^4r^4} + 
\frac{1}{2}B_{\rm y0} r \, , 
\eeqa
where $B_{\rm y0}$ is the magnitude of the external magnetic field which 
is chosen vertical, 
$k_{\rm r}$ and $k_{\rm y}$ are inverse length scales along the radial and 
vertical directions, respectively. The magnetic field lines, which follow 
from Eqs.~(\ref{eq:B_com_A}) and (\ref{eq:A-Psi-flux tube}) are displayed in 
Fig.~\ref{fig:mag_e} for $k_{\rm r}=k_{\rm y}=4$ Mm$^{-1}$. 
We choose and hold fixed 
the magnitude of the reference magnetic field $B_{\rm 0}$ in such way that 
the magnetic field within the flux tube, at $r=y=0$ Mm, is about 
$285$ Gauss, and $B_{\rm y0}\approx 11.4$ Gauss. For these values, the 
resulting magnetic field lines are predominantly vertical around 
the 
tube axis, 
$r=0$ Mm, while further out they are bent and $B_{\rm e}$ decays with 
distance. 
Such magnetic field lines correspond to an isolated axi-symmetric magnetic flux tube. 
It must be also noted that the magnetic field at 
the top of the simulation region is essentially uniform, with its value of $B_{\rm y0}$. 

Having specified the magnetic field, 
the equilibrium mass density and the gas pressure are evaluated from Eq.~(\ref{eq:force_free}) 
with the adaptation of 
the hydrostatic gas pressure $p_{\rm h}$ which is given by 
\beq
\label{eq:pres}
p_{\rm h}(y)=p_{\rm 0}~{\rm exp}\left[ -\int_{y_{\rm r}}^{y}\frac{dy^{'}}
{\Lambda (y^{'})} \right]\, ,
\eeq
%
where $y_{\rm r}=10$ Mm is the reference level, $p_{\rm 0}=0.01$ Pa is the gas pressure evaluated at $y=y_{\rm r}$, noting that
\begin{equation}
\Lambda(y) = \frac{k_{\rm B} T_{\rm h}(y)} {mg}
\end{equation}
%
is the pressure scale-height, and $T_{\rm h}(y)$ a hydrostatic 
temperature profile. 
We adopt $T_{\rm h}(y)$ for the solar atmosphere that is specified by the 
model developed by Avrett \& Loeser (2008). This temperature profile is 
smoothly extended into the corona. 
It should be further noted that in our model the solar photosphere occupies the region 
$0 < y < 0.5$ Mm, the chromosphere is sandwiched between $y=0.5$ Mm and the 
transition region located at $y\simeq 2.1$ Mm, and above this
height the atmospheric layers represent the solar corona. 

\section{Numerical simulations of MHD equations}\label{sec:num_sim_MHD}
%
%
%
In order to solve Eqs (\ref{eq:MHD_rho})-(\ref{eq:MHD_CLAP}) numerically, we use 
the FLASH code (Fryxell et al. 2000; 
Lee \& Deane 2009; Lee 2013),
in which a third-order unsplit Godunov-type solver with various slope limiters 
and Riemann solvers as well as Adaptive Mesh Refinement (AMR) 
(MacNeice et al. 1999) are implemented. The minmod slope limiter and the Roe Riemann 
solver (e.g., T\'oth 2000) are used. We set the simulation box as
$(-1.5\, {\rm Mm},1.5\, {\rm Mm}) \times (0\, {\rm Mm},18\, {\rm Mm}) \times (-1.5\, {\rm Mm},1.5\, {\rm Mm})$ 
and impose fixed in time boundary conditions for all plasma quantities
in 
all directions. 
%
In our present work, we use a static, non-uniform grid with a minimum (maximum) level 
of refinement set to $2$ ($5$). 
Note that small size blocks of numerical grid occupy 
the level up to $y=4$ Mm (Fig.~\ref{fig:blk}), 
and every numerical block consists of $8\times 8\times 8$ identical 
numerical cells. This results in the smallest spatial resolution of $23.4$ km 
below the level $y=4$ Mm 
and allows to 
well resolve 
spatial structures 
in the low solar corona. 
%
\subsection{Initial perturbations}
%
%
%
The atmospheric magnetic field is continuously disturbed by large-scale dynamical perturbations that are able to transfer kinetic energy 
({\it e.g.} buffeting due to the granular motion in the photosphere, or flare-driven blast waves in the upper atmosphere, reconnection-driven shocks, etc.). 
We assume that the instigator of changes in the magnetic field has such a nature. However, its exact property 
is not specified, and initially (at $t=0$ s) we perturb the above described equilibrium impulsively by a Gaussian pulse 
in the azimuthal component of velocity, $V_{\rm \theta}$, viz.,
\beq\label{eq:perturb}
V_{\theta}
= A_{\rm v} \frac{{\tilde r}}{w}
\exp\left[ -\frac{{\tilde r}^2 + (y-y_{\rm 0})^2}{w^2} \right]\, ,
\eeq
where ${\tilde r}^2=(x-x_{\rm 0})^2+z^2$ is the squared radius, 
$A_{\rm v}$ is the amplitude of the pulse, 
$(x_{\rm 0}, y_{\rm 0},0)$ its position, 
and $w$ its width. We set $A_{\rm v}=6$ km s$^{-1}$, $w=150$ km, and 
$y_{\rm 0}=500$ km, and hold them fixed, while we allow $x_{\rm 0}$ to attain one of these two values: (a) $x_{\rm 0}=0$ km; (b) $x_{\rm 0}=100$ km. 
This value of $A_{\rm v}$ results in 
the effective maximum velocity of about $2.56$ km s$^{-1}$ (Fig.~\ref{fig:blk}, color map), 
and the value of $y_{\rm 0}$ shows that 
the system is perturbed 
right 
at the top of the photosphere. 

Note that in our 3D model, the torsional Alfv\'en waves decouple linearly from magnetoacoustic 
waves. They can be described solely by $V_{\rm \theta}(x,y,t)$ (see Murawski et al. 2015). 
As a result, the initial pulse triggers Alfv\'en waves. 

%
\section{Results of numerical simulations}\label{sec:resultsOFnumSIM}
%
We simulate small amplitude and impulsively excited Alfv\'en waves and investigate
their propagation along the magnetic field lines 
which are associated with the flux tube (Fig.~\ref{fig:mag_e}). 
We consider the following two cases: (a) centrally-launched initial pulse; (b) off-centrally-launched initial pulse, 
and describe the corresponding results in the following parts of the paper.  
\subsection{Centrally launched pulse}
The case of a centrally-launched pulse corresponds to the location of the initial pulse at the flux tube axis, which is realized by setting 
$x_{\rm 0}=0$ Mm in Eq.~(\ref{eq:perturb}). 
\begin{figure}
\centering{
           \includegraphics[scale=0.2,angle=0]{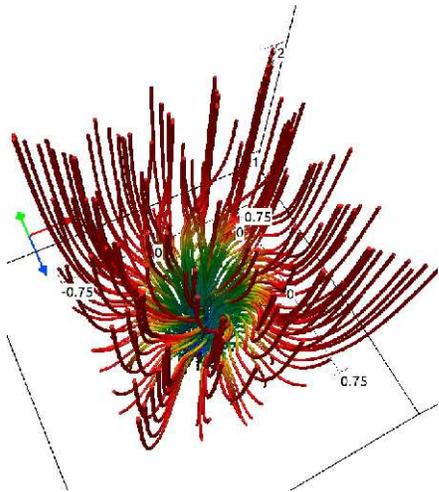}
          }
\caption{\small Spatial profiles of magnetic field lines
         at 
$t=200$ s for the case of centrally-launched initial pulse. 
        }
\label{fig:B-lines}
\end{figure}
%
A 3D view of the 
magnetic field lines of the flux tube 
is shown at $t=200$ s in Fig.~\ref{fig:B-lines}. 
It is clear that as a result of torsional Alfv\'en propagation, magnetic field lines are twisted and this twist 
is localized essentially below the transition region. 

%
\begin{figure}
\centering{
           \includegraphics[scale=0.225,angle=0]{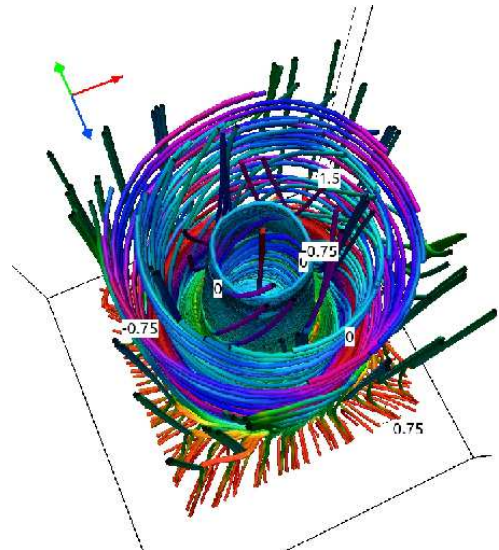}
           \includegraphics[scale=0.225,angle=0]{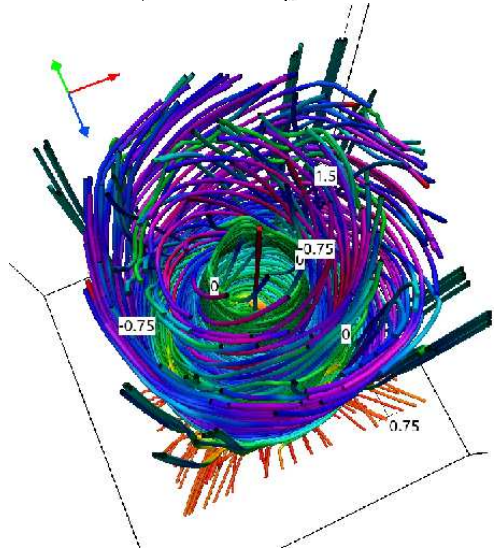}
          }
\caption{\small Spatial profiles of streamlines 
         at $t=290$ s (top)
and $t=420$ s (bottom) for the case of centrally-launched initial pulse. 
        }
\label{fig:V-lines}
\end{figure}
%
Figure~\ref{fig:V-lines} presents velocity streamlines which exhibit the twisters' motion. 
At $t=290$ s we discern a centrally located shell 
in which plasma counter-rotates in comparison to the rotation of the external shell (the top panel). 
At the later moment of time, this central shell is more developed (the bottom panel). 
It is to be further noted that these concentric magnetic shells have their dimension of the order of $120$ km. 

%
\begin{figure}
\centering{
           \includegraphics[scale=0.275,angle=0]{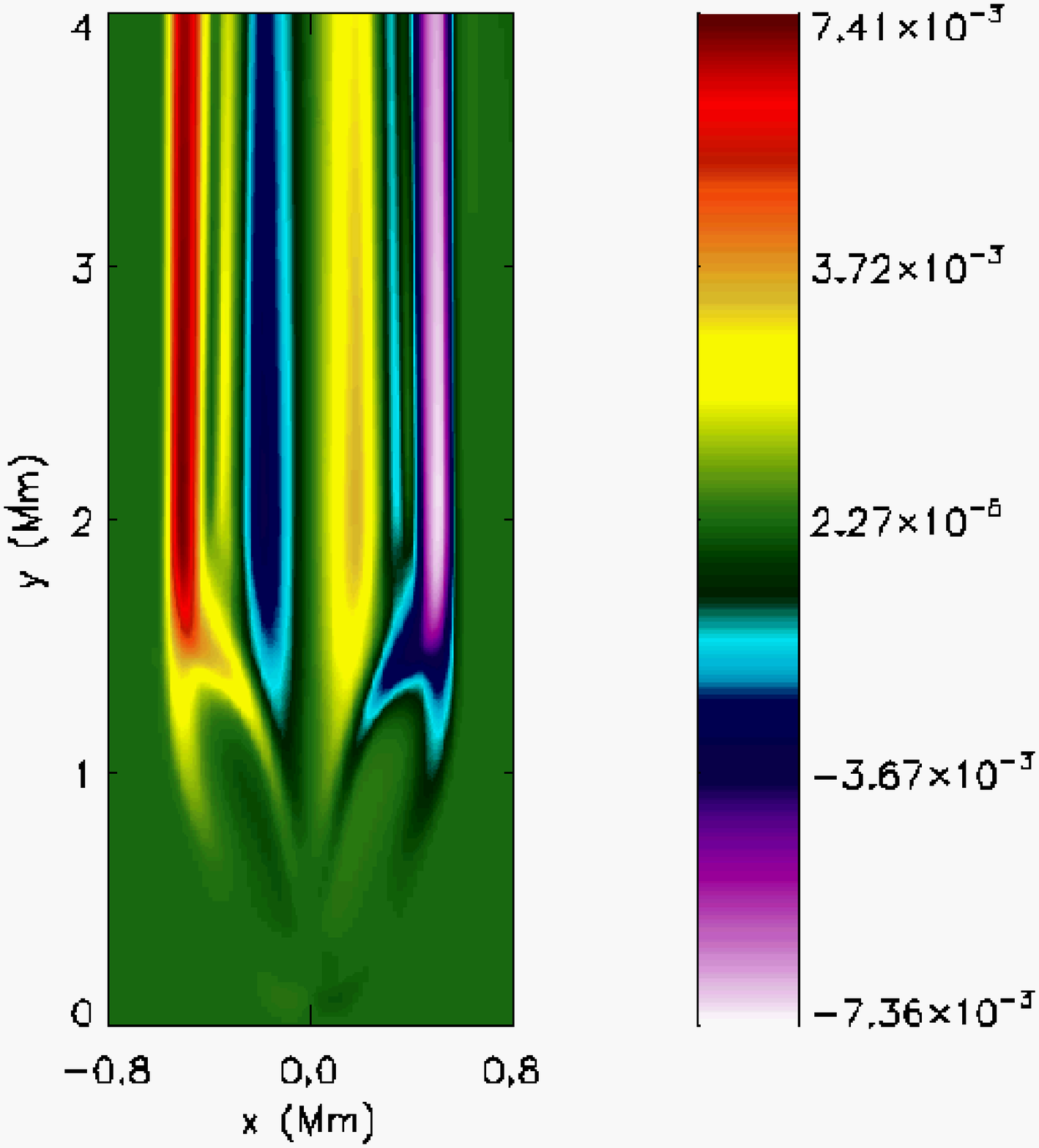}\\
\vspace{-0.75cm}
           \includegraphics[scale=0.275,angle=0]{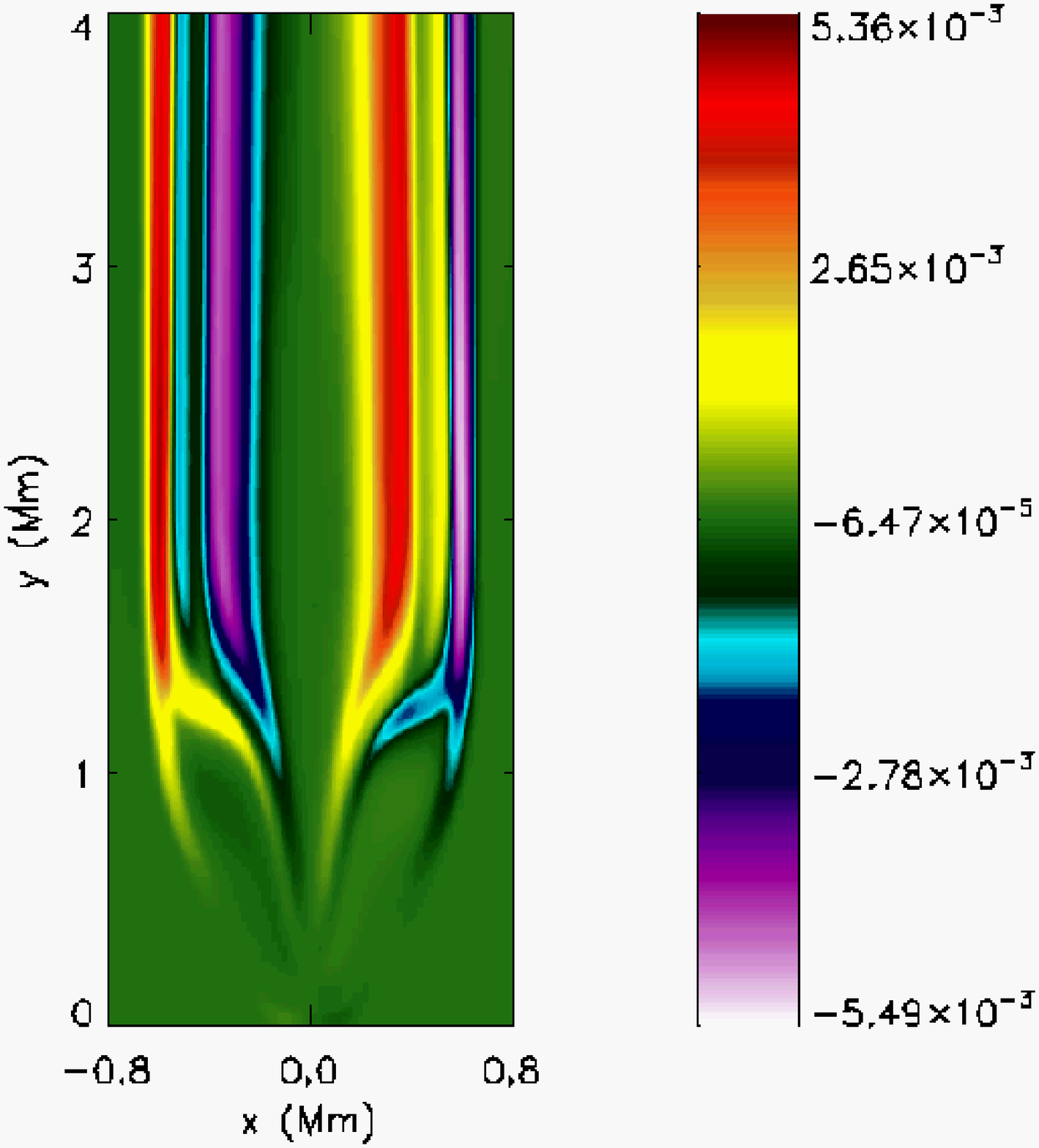}
          }
\caption{\small Spatial profiles of transverse velocity $V_{\rm z}(x,y,z=0,t)$
         at 
$t=290$ s (top) 
and $t=420$ s (bottom) for the case of centrally-launched pulse, $x_{\rm 0}=0$ km.
        }
\label{fig:Vz}
\end{figure}
%
Figure~\ref{fig:Vz} shows the 
vertical cross-sections of 
$V_{\rm z}(x,y,z=0)=V_{\rm \theta}(x,y,z=0)$
exhibiting the origin of various magnetic interfaces (multi-shells) at 
an upper chromospheric height of $y=1.9$ Mm, which is located about $200$ km below the transition region. 
It is to be noted that the magnetic field perturbations are confined to atmospheric layers located below the solar transition region 
(Murawski and Musielak 2010; Murawski et al. 2015), 
while the velocity perturbations easily penetrate into the corona and transfer the available mechanical energy (Fig.~\ref{fig:Vz}). 
The 
vertical cut at the upper chromospheric height in the velocity 
clearly shows the two oppositely oriented shells, 
where plasma is counter-streaming to each other and forming two twisting plasma shells. 
As a result, a distinct set of magnetic interfaces is formed, where Alfv\'en wave velocity perturbations are also generated.


%
\begin{figure*}
\centering{
           \includegraphics[scale=0.275,angle=0]{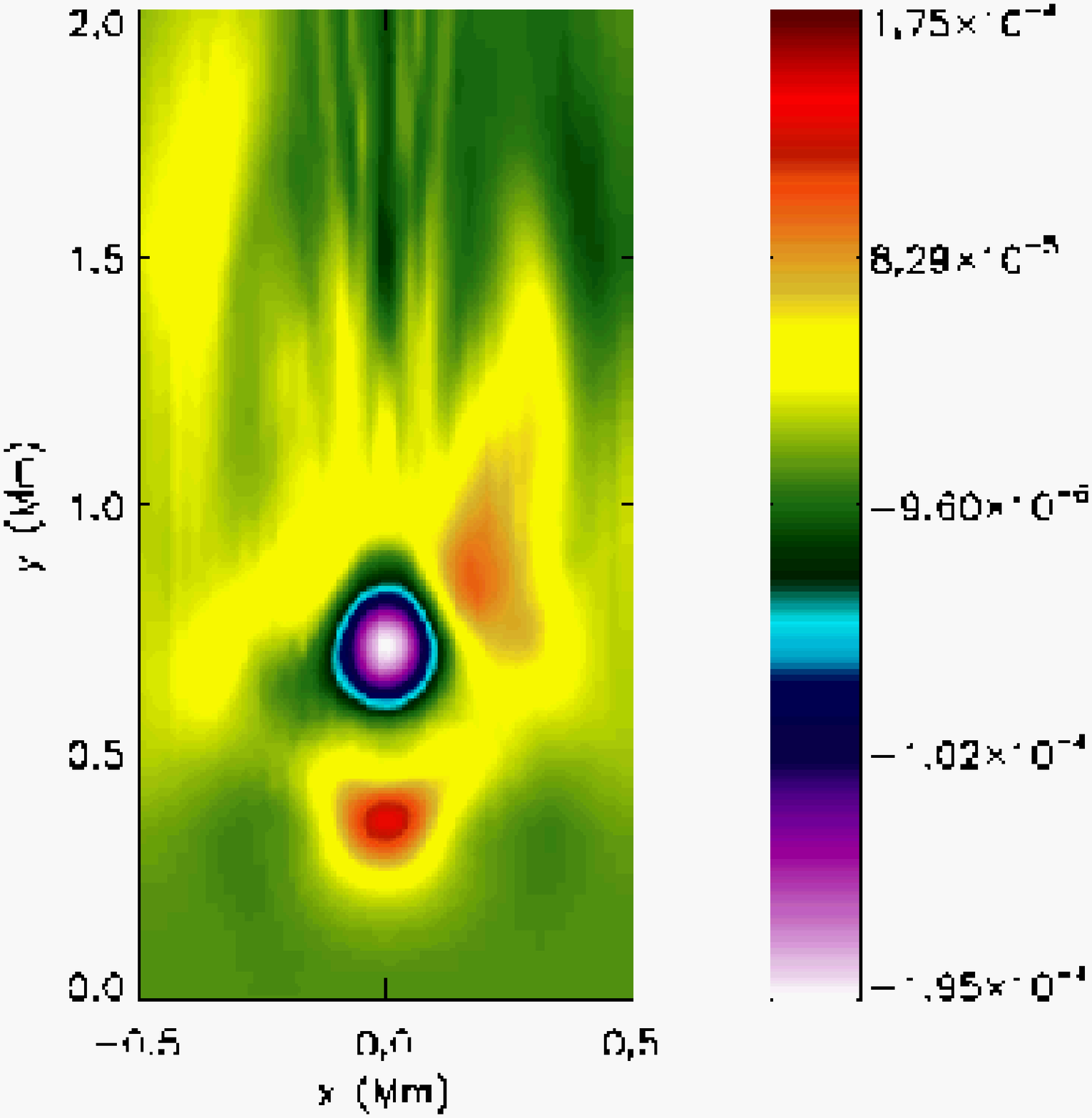}
           \includegraphics[scale=0.275,angle=0]{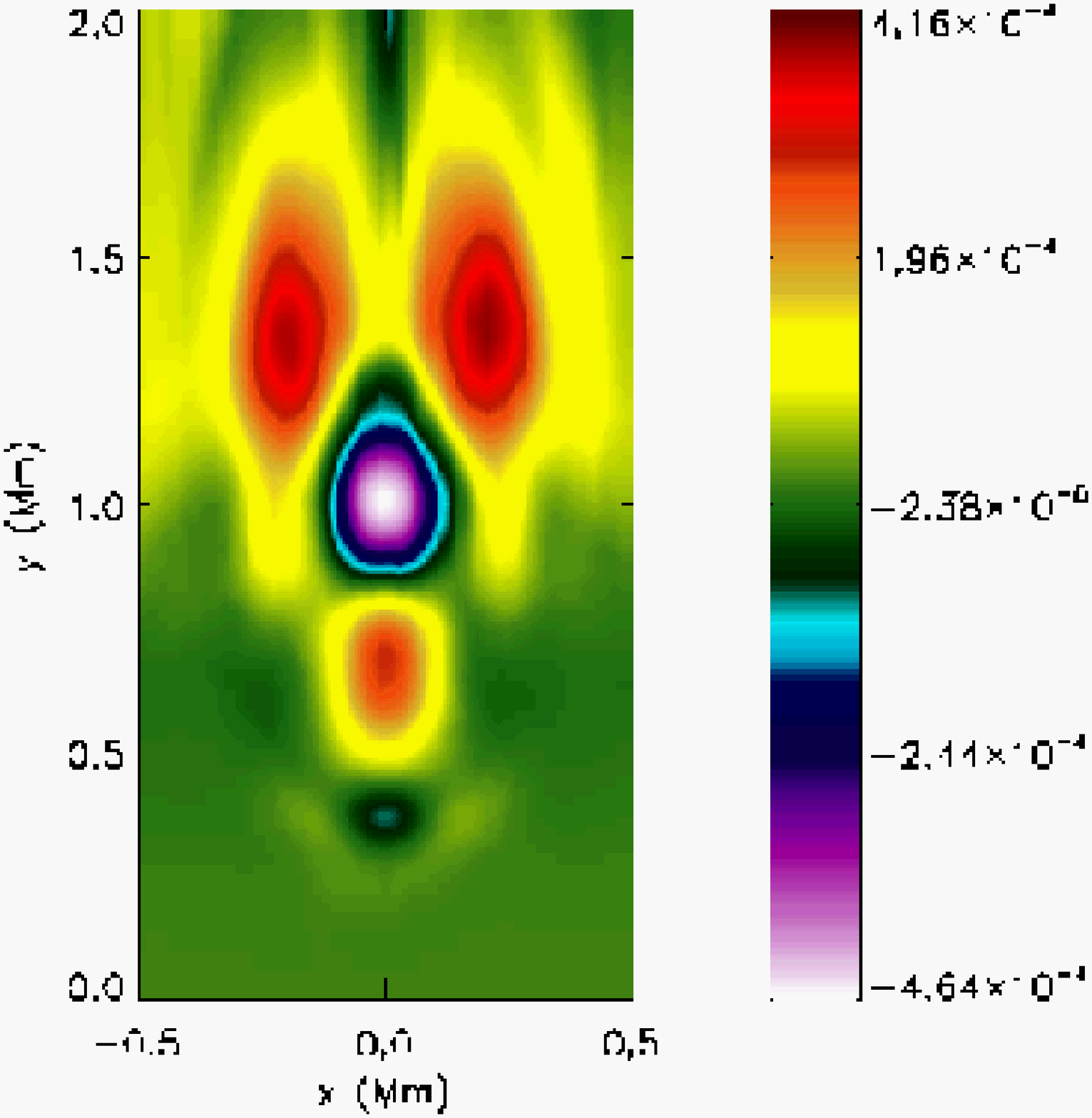}\\
           \includegraphics[scale=0.275,angle=0]{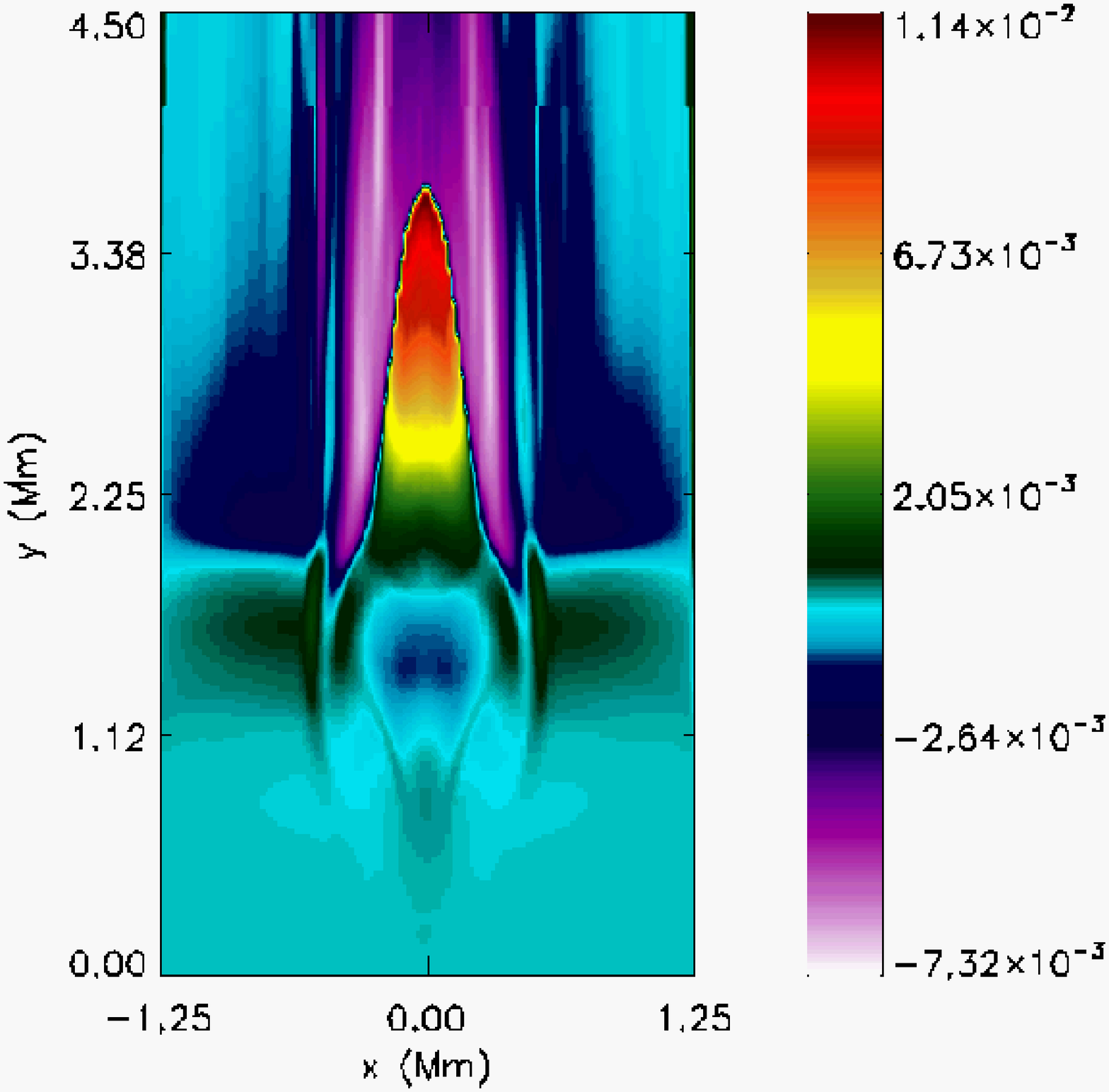}
           \includegraphics[scale=0.275,angle=0]{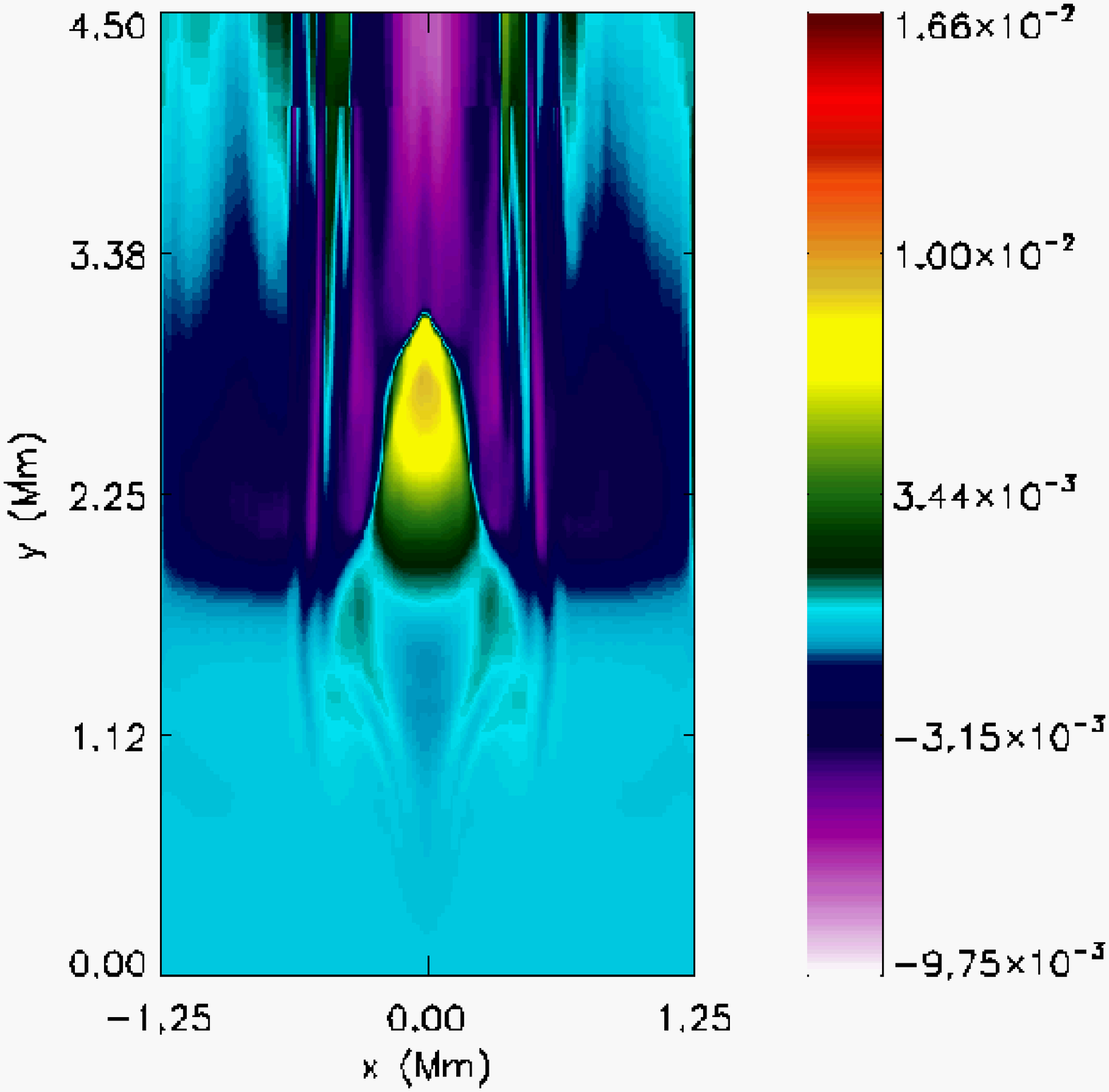}
          }
\caption{\small Spatial profiles of 
 $V_{\rm y}(x,y,z=0,t)$
         at 
         $t=50$ s, 
         $t=100$ s, 
         $t=290$ s 
and 
         $t=420$ s 
         (from top-left to bottom-right) 
for the case of centrally-launched pulse, $x_{\rm 0}=0$ Mm. 
        }
\label{fig:Vy}
\end{figure*}
%
Figure~\ref{fig:Vy} illustrates spatial profiles of the vertical component of velocity, $V_{\rm y}(x,y,z=0,t)$, 
drawn in the vertical plane for $z=0$. 
As a result of the initial pulse in $V_{\rm \theta}$, it follows from the Bernoulli's law that the under-pressure region 
is generated shortly after 
$t=0$ s 
and the ambient plasma is sucked into the region where the initial pulse was launched from. 
This under-pressure sucks the ambient plasma, resulting in a converging into the launching region 
flow. Such flow is well seen at $t=50$ s (the top-left panel). 
Note the down-flowing plasma in the patch located at $(x=0,y\approx 0.75)$ Mm, 
while the up-flowing gas is located at $(x=0,y\approx 0.3)$ Mm. 
Later on, at $t=100$ s these patches move upward and new regions of up-flowing plasma are generated at $(x\approx \pm 0.2,y\approx 1.4)$ Mm (top-right).
Some of the up-flowing plasma is attracted by the gravity force, resulting in 
a down-flowing blob of the plasma velocity of $V_{\rm y}\approx -6$ km s$^{-1}$, 
which is seen at $t=290$ s above the transition region in the form of the violet map (the bottom-left panel). 
This free-falling plasma clashes with the first jet which originates right below the transition region (the red patch in the bottom-left panel). 
As a result of this clash a shock is generated. The second jet is seen at $t=420$ s (bottom-right). 


%
\begin{figure}
\centering{
           \includegraphics[scale=0.275,angle=0]{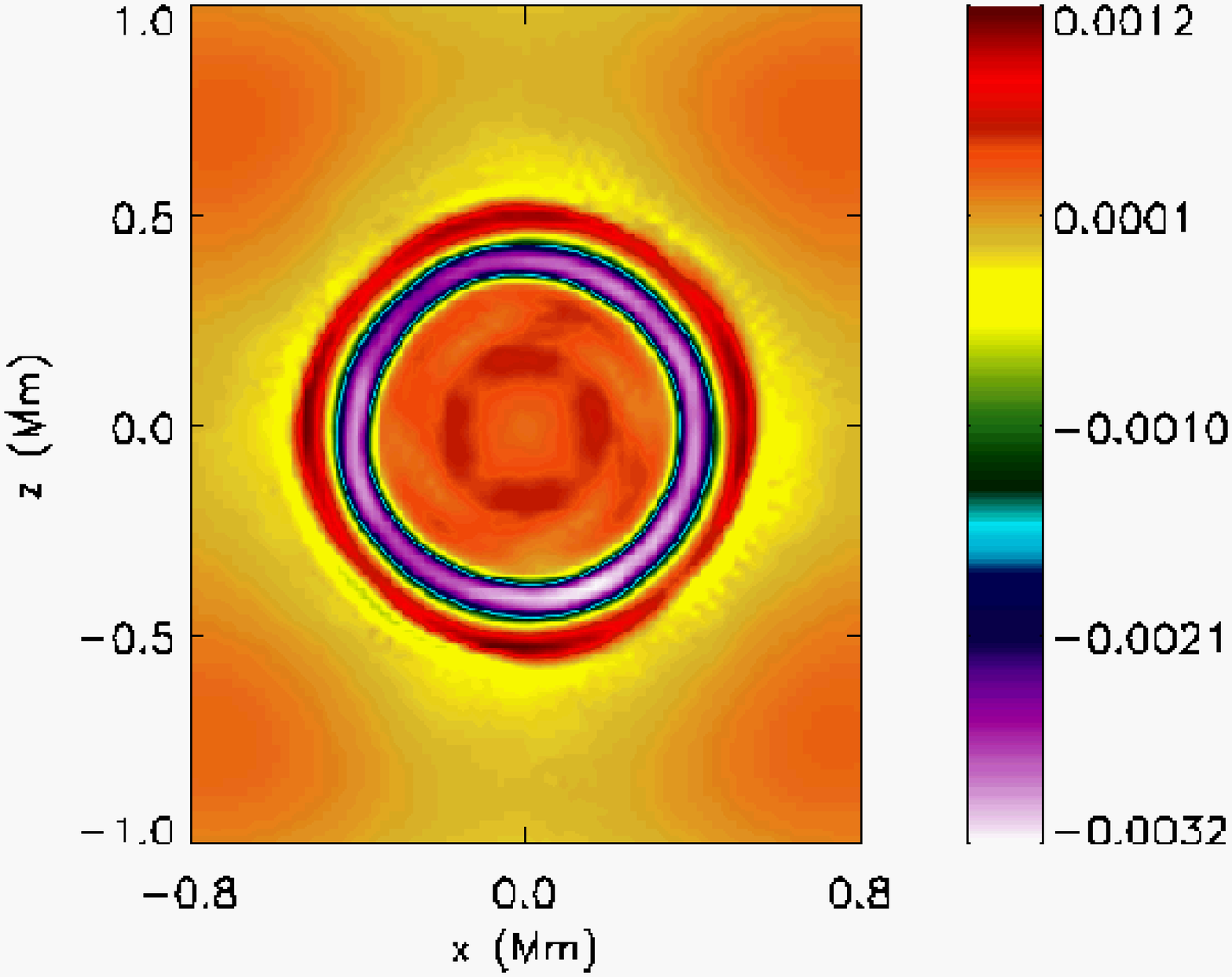}
           \vspace{-1.5cm}
           \includegraphics[scale=0.275,angle=0]{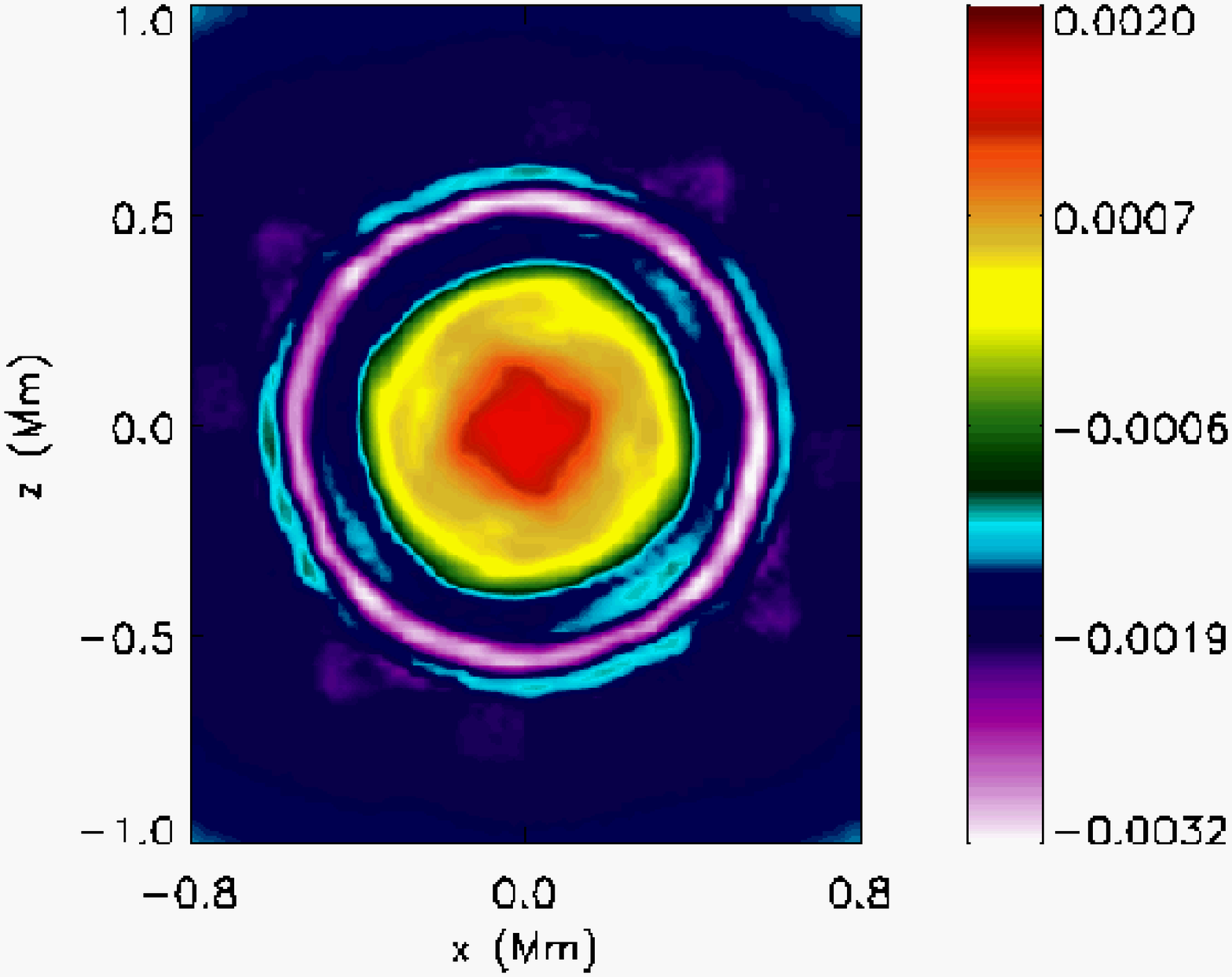}
          }
\caption{\small Spatial profiles of 
 $V_{\rm y}(x,y=1.9,z,t)$
         at 
         $t=290$ s (top) 
and 
         $t=420$ s (bottom) for the case of the centrally-launched initial pulse, $x_{\rm 0}=0$ km. 
        }
\label{fig:Vy_y=1.9}
\end{figure}
%

The vertical components of velocity, $V_{\rm y}$, are displayed in the horizontal plane just below the transition region, at $y=1.9$ Mm, 
in Fig.~\ref{fig:Vy_y=1.9}. The development of concentric shells is well seen at $t=290$ s (top); 
the inner shell is represented by the violet ring with down-flowing plasma with its speed $V_{\rm y}\approx 3$ km s$^{-1}$, 
while in the outer shell, marked by the red ring, and the plasma flows upward with its speed $V_{\rm y}\approx 1$ km s$^{-1}$. 
In the center the plasma begins to flow upwards. At $t=420$ s this central flow develops the structure with $4$ discernible arms (bottom). 
Note that the vertical flow is out of phase to $V_{\rm z}$ at each instance in a particular shell at a given chromospheric height (compare with Fig.~\ref{fig:Vz}). 
This is a unique correlation of periodic changes in the vertical flows (i.e., up- and down-flows) 
negatively coupled with the Alfv\'en wave velocity perturbations in a chromospheric shell. 

%
\begin{figure}
\centering{
           \includegraphics[scale=0.275,angle=0]{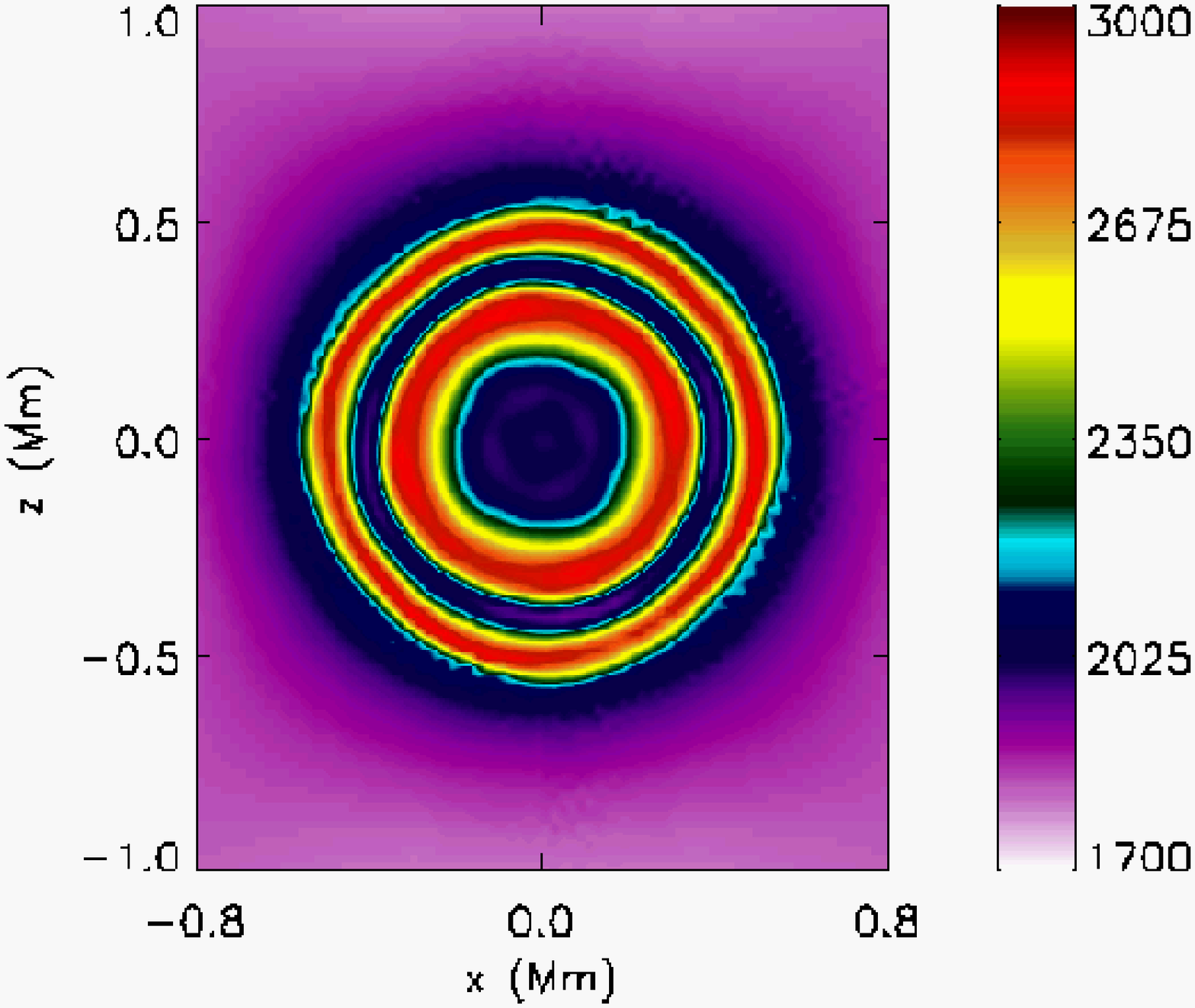}
           \vspace{-1.5cm}
           \includegraphics[scale=0.275,angle=0]{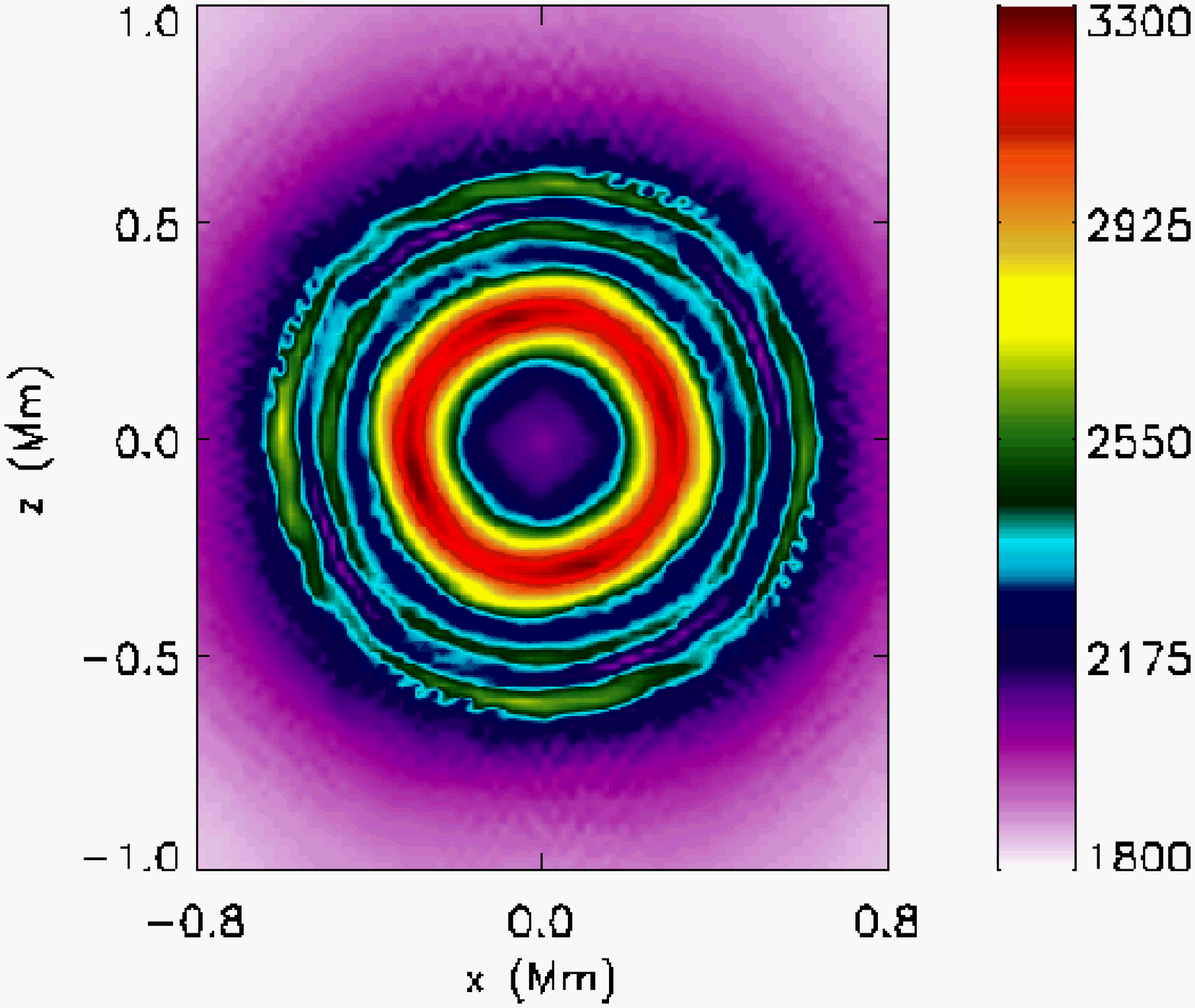}
          }
\caption{\small Spatial profiles of mass density $\varrho(x,y=1.9,z,t)$
         at $t=290$ s (top) and 
$t=420$ s (bottom) for the case of centrally-launched initial pulse. 
        }
\label{fig:rho-x-y}
\end{figure}
%
The multi-shells can also be traced on the horizontal profiles of mass density, $\varrho(x,y=1.9,z,t)$, which are displayed in Fig.~\ref{fig:rho-x-y}. 
The initial profile (not shown) is altered at $t=290$ s with $2$ well-developed dense concentric shells (top). 
At a later moment of time, $t=420$ s, there are more shells seen (bottom). 
These shells result from up-flowing and down-flowing plasma and from the magnetoacoustic-gravity waves, 
which propagate in the horizontal direction. 
These magnetoacoustic waves are surface-gravity-like waves and they are driven by the oscillations in $V_y$.
These waves propagate along the transition region. 
The propagating upward plasma brings a dense gas from lower atmospheric regions, 
while down-falling flow leads to plasma evacuation, which is well seen in the rings. 
Within the center of the flux-tube, the oscillations in $V_y(x,y=1.9,z)$ and $\varrho(x,y=1.9,z)$ are in anti-phase. 
Compare Figs.~\ref{fig:Vy_y=1.9} and ~\ref{fig:rho-x-y}. 

%
\begin{figure}
\centering{
           \includegraphics[width=6.2cm,angle=0]{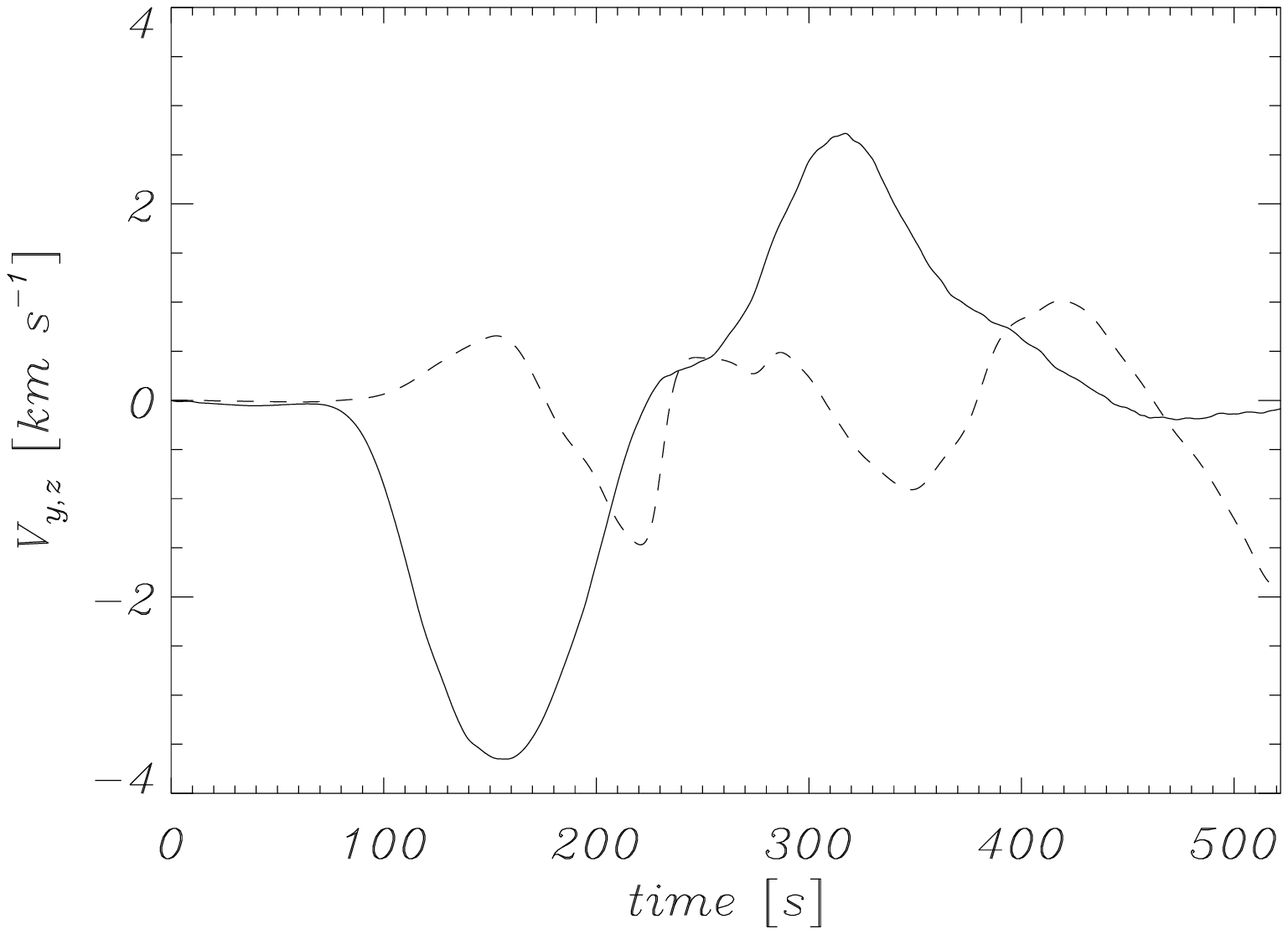}\\
           \includegraphics[width=6.2cm,angle=0]{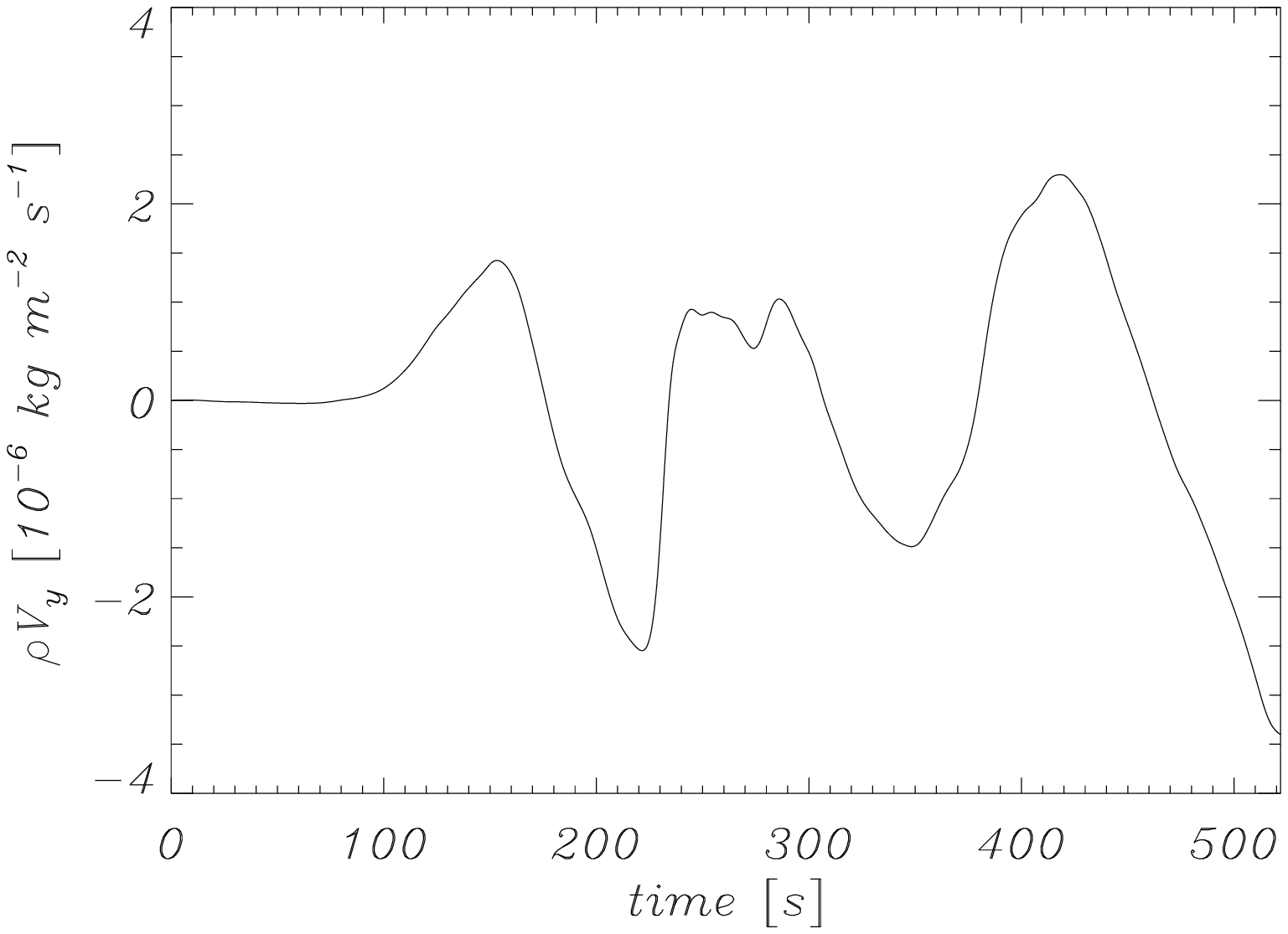}\\
           \includegraphics[width=6.2cm,angle=0]{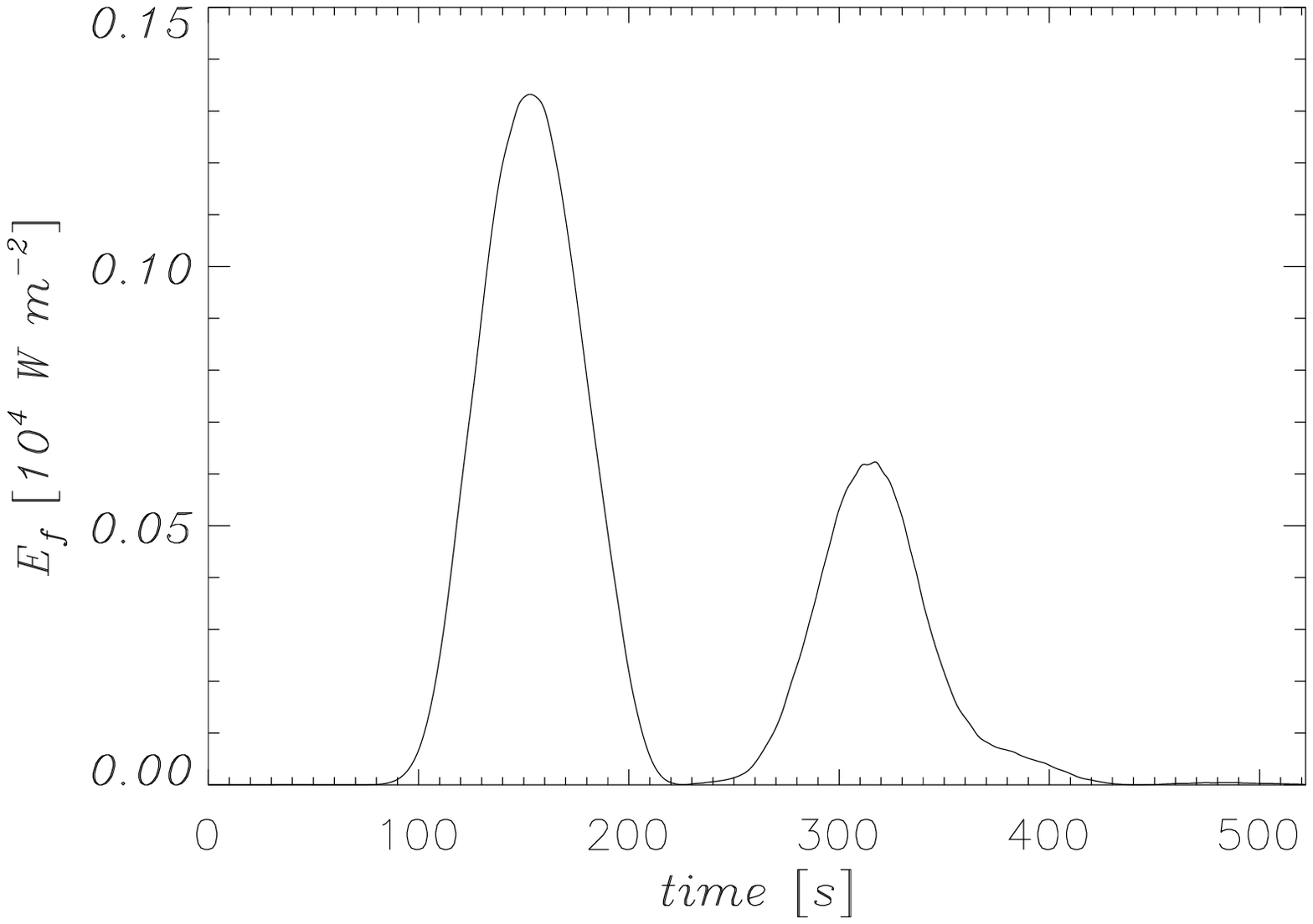}\\
          }
\caption{\small
         Time-signatures of $V_{\rm z}$ (top, solid line), 
         $V_{\rm y}$ (top, dashed line), vertical component of the mass density flux 
         $\varrho V_{\rm y}$ (middle), 
         and approximated Alfv\'en waves energy flux (bottom), evaluated at 
         the detection point ($x=0.1$ Mm, $y=1.9$ Mm, $z=0.1$ Mm) for the case of $x_{\rm 0}=0$ km. 
        }
\label{fig:ts}
\end{figure}
%
Figure~\ref{fig:ts} (top) displays time-signatures of the transversal ($V_{\rm z}$) and vertical ($V_{\rm y}$) 
components of velocity, represented respectively by solid and dashed lines, and collected 
at the detection point which is located at ($x=0.1$ Mm, $y=1.9$ Mm, $z=0.1$ Mm). 
These time-signatures reveal perturbations in which $V_{\rm z}$ and $V_{\rm y}$ are approximately in anti-phase. 
Hence, we infer that up-flows (which correspond to $V_{\rm y}>0$) lead to clock-wise oscillations in $V_{\rm \theta}$ 
(represented in the vertical plane by $V_{\rm z}$), 
while down-flows are associated with anti-clock-wise oscillations in $V_{\rm \theta}$. 
The Alfv\'en wave energy flux can be approximated by the following relation (Vigeesh et al. 2012): 
\beqa\label{eq:E_flux}
Ef \approx \varrho_{\rm e} V_{\rm \theta}^2 c_{\rm A}\, ,
\eeqa
where $\varrho_{\rm e}$ is the unperturbed mass density, $V_{\rm \theta}$ is the azimuthal component of the Alfv\'en speed 
and $c_{\rm A}$ is the Alfv\'en speed  
which is expressed as 
%
\beq
\label{eq:ca}
c_{\rm A}^2(x,y,z) = \frac{ {B}^2_{\rm e}(x,y,z) }{{\mu \varrho_{\rm e}(x,y,z)}}\, .
\eeq
Note that the vertical flows associated with the magnetic twister carry the mass-flux of its maximum of 
$6\times 10^{-6}$ kg m$^{-2}$ s$^{-1}$ to the inner coronal heights, and thereby they also contribute to the mass supply 
in the region of nascent solar wind formation (Figure~\ref{fig:ts}, middle). 
The net energy flux carried by these twisters is about $0.2\times 10^4$ W m$^{-2}$ (Fig.~\ref{fig:ts}, bottom), which 
is almost sufficient for the required coronal heating of the quiet-Sun and the solar wind energy losses. 
\subsection{Off-centrally-launched pulse}
We now consider the case of off-centrally-launched pulse that
corresponds to the location of the initial pulse out off the flux tube axis, which is realized by setting 
$x_{\rm 0}=0.1$ Mm (Eq.~\ref{eq:perturb}).
%
\begin{figure}
\centering{
           \includegraphics[scale=0.275,angle=0]{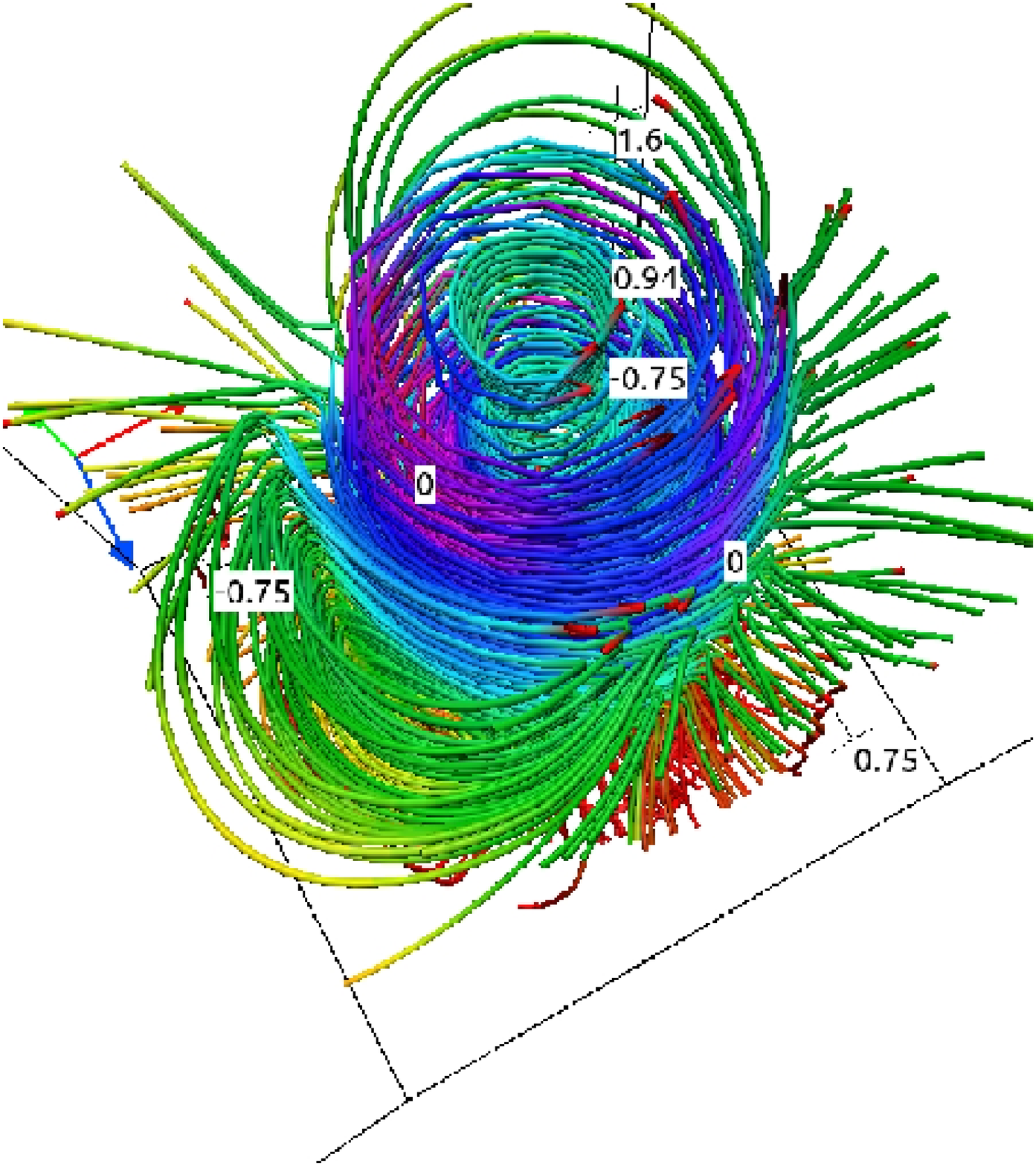}
           \includegraphics[scale=0.275,angle=0]{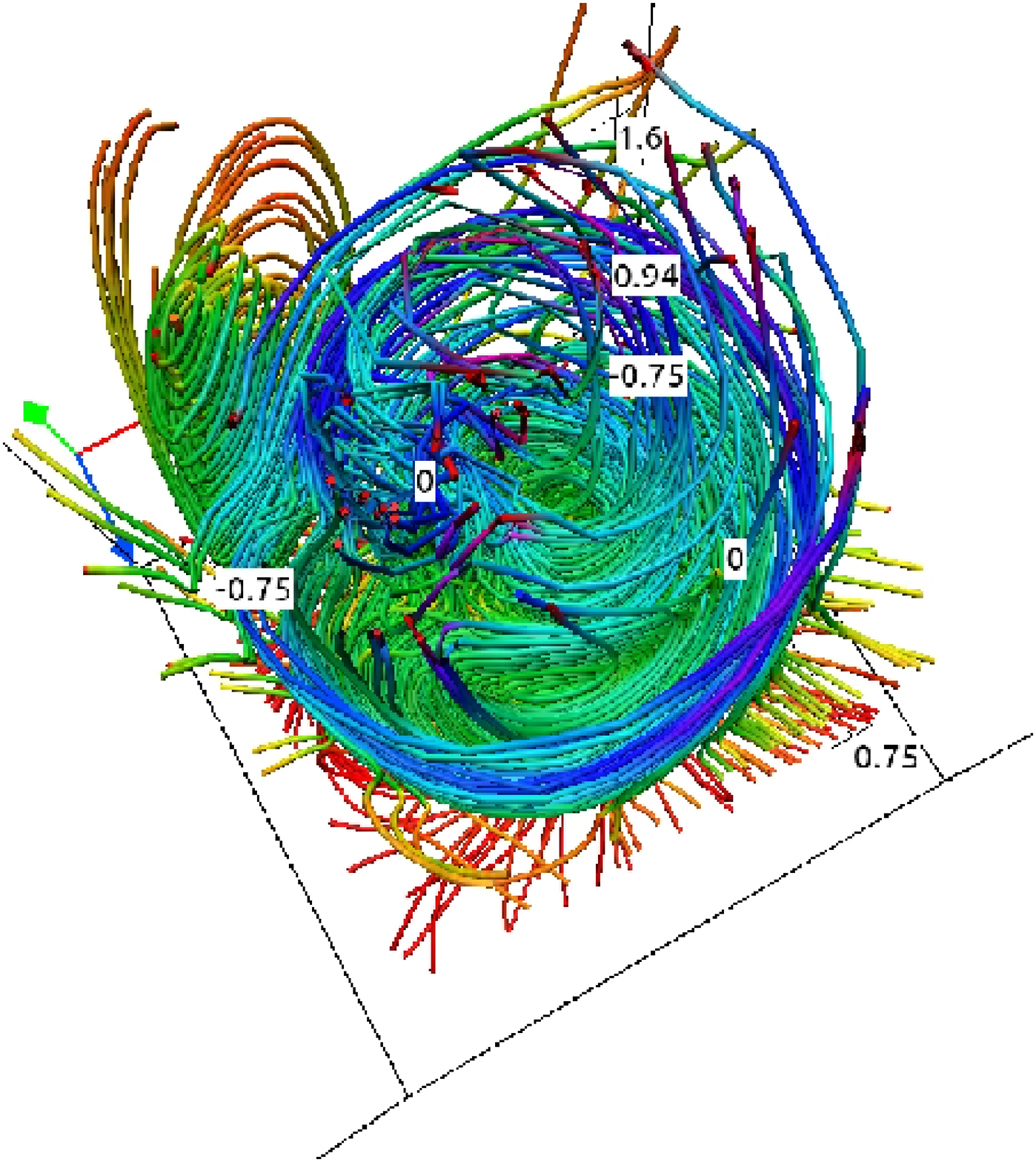}
          }
\caption{\small Spatial profiles of streamlines 
         at $t=150$ s (top)
and $t=300$ s (bottom) for the case of $x_{\rm 0}=100$ km. 
        }
\label{fig:V-lines-off}
\end{figure}
%
%
\begin{figure}
\centering{
           \includegraphics[scale=0.275,angle=0]{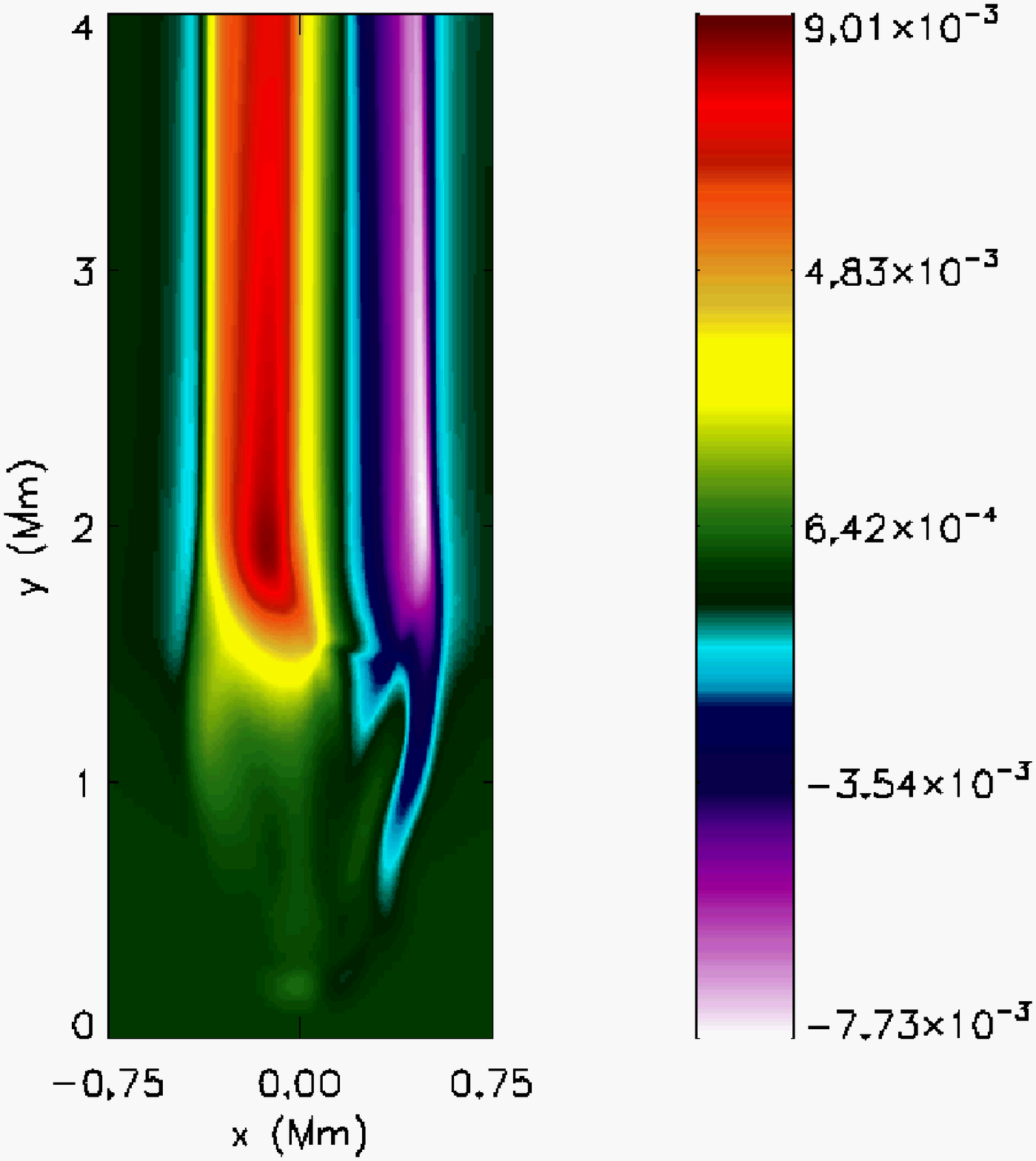}\\
           \includegraphics[scale=0.275,angle=0]{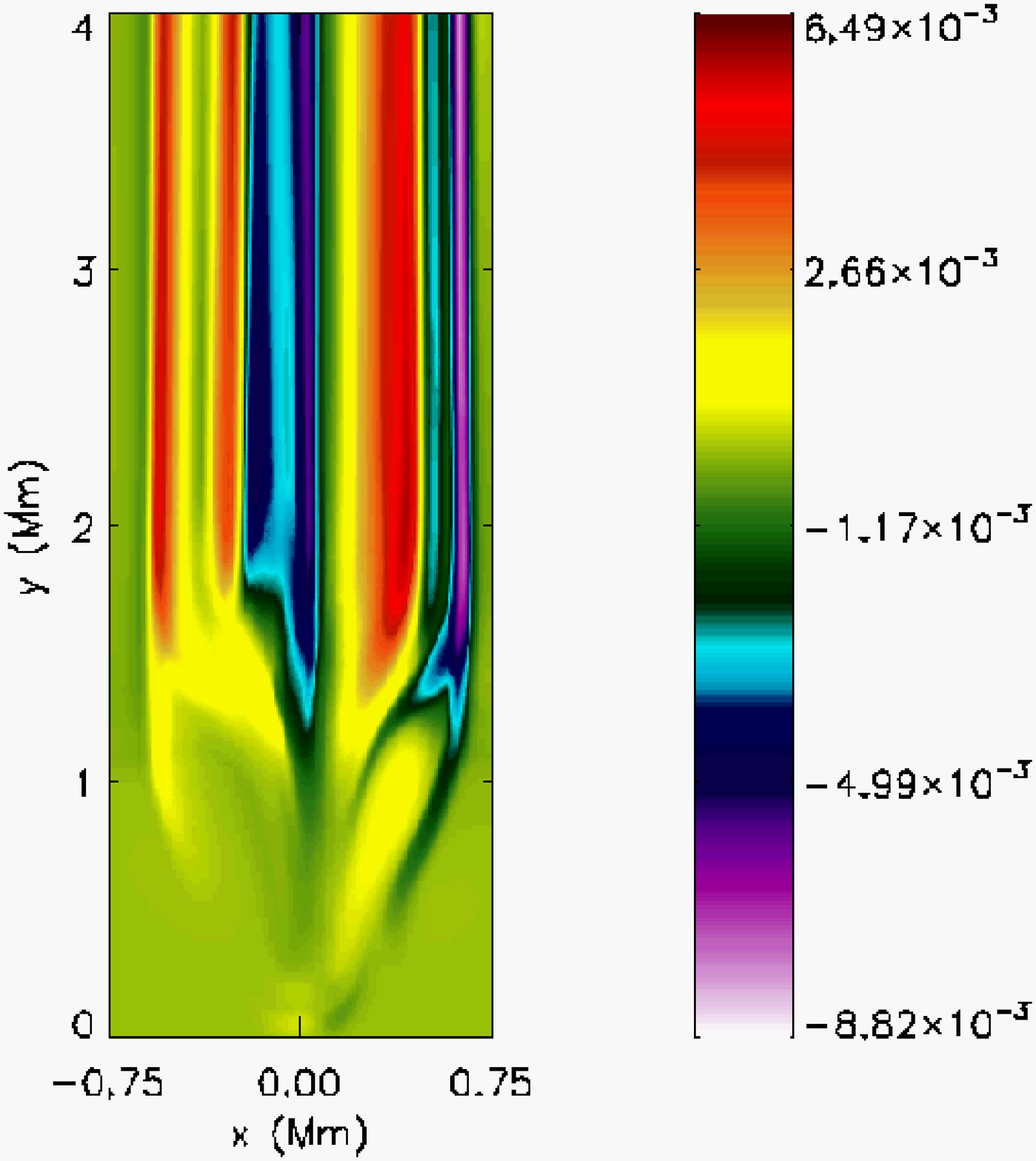}
          }
\caption{\small Spatial profiles of transverse velocity $V_{\rm z}(x,y,z=0,t)$
         at 
$t=150$ s (top) 
and $t=300$ s (bottom) for the case of $x_{\rm 0}=100$ km. 
        }
\label{fig:Vz-off}
\end{figure}
%
%
\begin{figure}
\centering{
           \includegraphics[scale=0.275,angle=0]{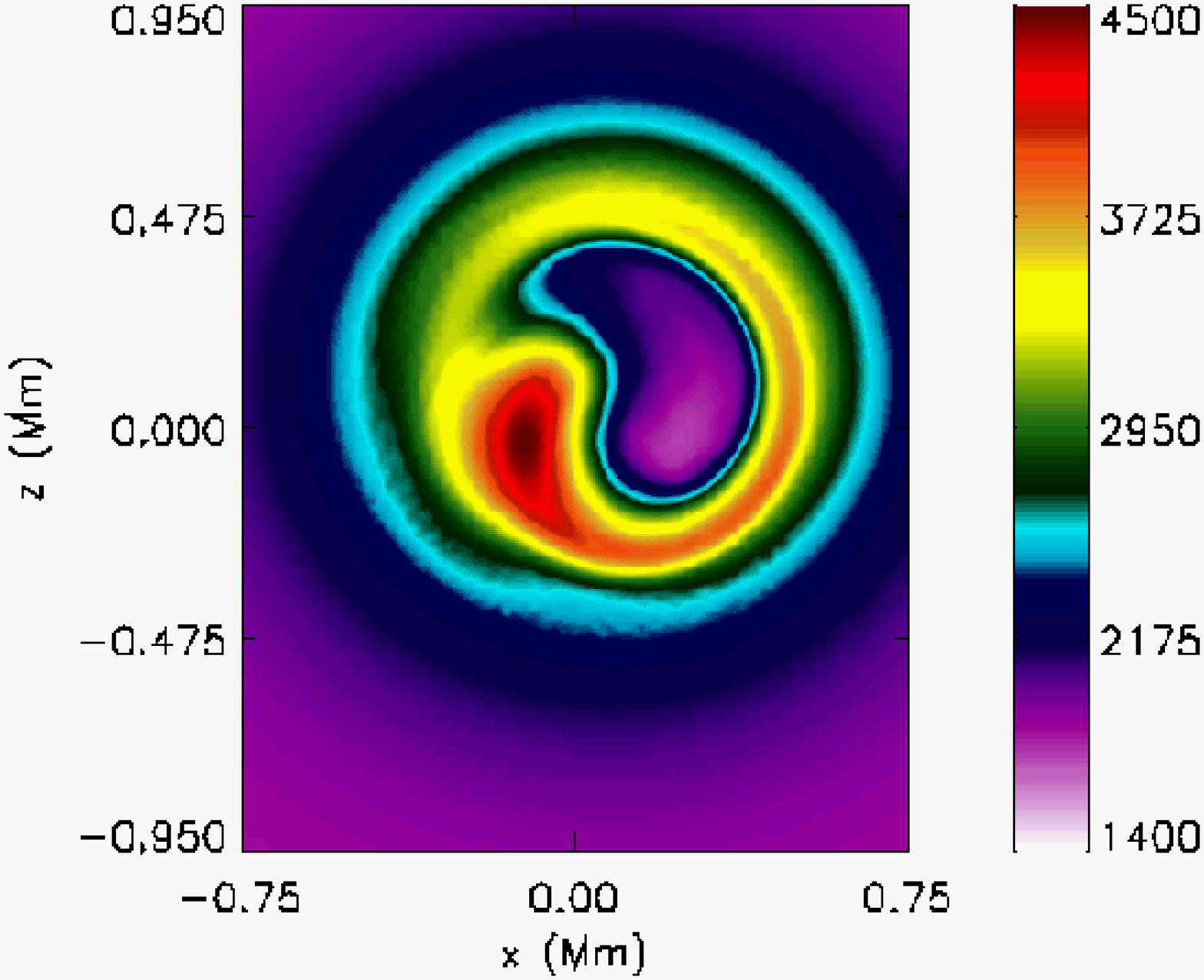}
            \vspace{-1.5cm}
           \includegraphics[scale=0.275,angle=0]{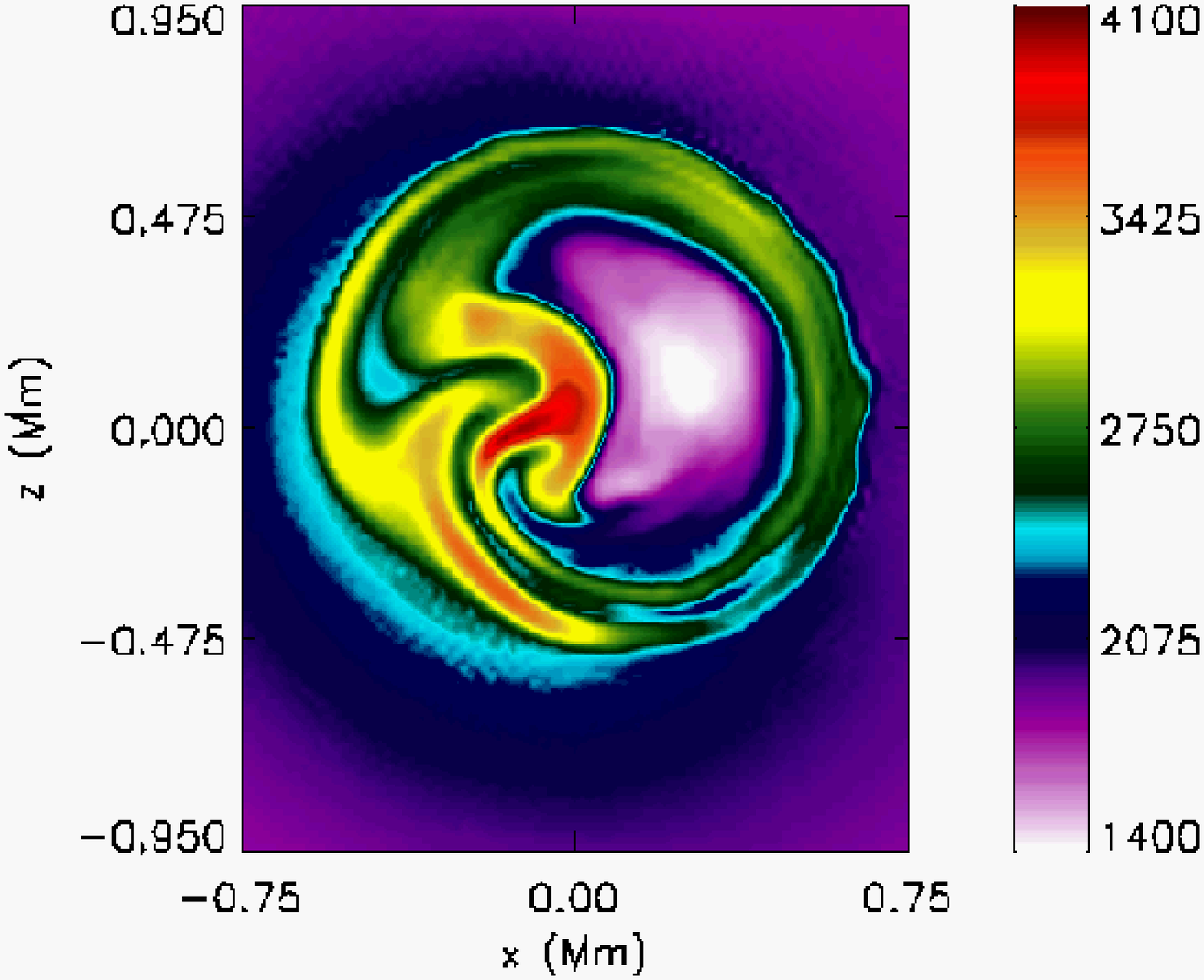}
          }
\caption{\small Spatial profiles of mass density $\varrho(x,y=1.9,z,t)$
         at $t=150$ s (top) 
and $t=300$ s (bottom) for $x_{\rm 0}=0.1$ Mm.  
        }
\label{fig:rho-x-y-off}
\end{figure}
%
The spatial profiles of streamlines are drawn in Fig.~\ref{fig:V-lines-off}. 
At $t=150$ s, we observe a development of the inner and outer eddies, as well as side vortices, 
seen at the bottom-left and top-right corners (top). 
At a later moment of time, $t=300$ s the flow pattern is more complex with more local eddies 
generated as a result of energy cascade into smaller scales (bottom). 
By comparing this with Fig.~\ref{fig:V-lines}, we conclude that the off-central pulse results in a more complex flow 
than the central pulse. 
This conclusion is further supported by the results of Fig.~\ref{fig:Vz-off} that illustrates 
vertical profiles of $V_{\rm z}(x,y,z=0)$ 
and Fig.~\ref{fig:rho-x-y-off} which shows the horizontal profiles of $\varrho(x,y=1.9,z)$. 
The symmetry of the centrally-launched pulse (Fig.~\ref{fig:Vz}) 
is broken for the case of the off-centrally-launched pulse (Fig.~\ref{fig:Vz-off}), 
with complex structures developed in the latter case. 
%
%
%
\section{Summary and Conclusions}\label{sec:Summary}
%

In this paper, we simulated impulsively generated, either centrally or off-centrally, Alfv\'en waves in a gravitationally stratified 
and magnetically confined solar magnetic flux tube with the temperature profile 
of Avrett and Loeser (2008). 
A novel result is that an outer shell with Alfv\'en wave velocity perturbation is generated within the flux tube at small spatial scales. 
The plasma associated with this shell rotates following the magnetic field lines at this interface with the velocity streamlines forming the swirls. 
The vertical flows are associated with this dynamics and they feedback energy to generate one more inner shell with Alfv\'en wave velocity perturbation 
within the tube. This process repeats in time resulting in a third inner shell. 
On a time-scale of few hundred seconds, 
the multi-shell magnetic twister is fully developed in the tube. 
When we view it from top of the tube, the collective appearance shows up as the perturbations in the magnetic field lines 
with the streamline rotational motion constituting the multi-shell magnetic twister. 
Each distinct appearance of the multiple plasma shells 
grows in time as every shell is demarcated by the transition in density, 
which results in the emission $I \sim N_{\rm e}^2$. 
The scenario of multi-shells development is more complex in the case of the off-centrally launched Alfv\'en pulse, 
with a larger number of asymmetric shells generated in the system. 

The main new result is that the multi-shell magnetic twister and 
associated multiple torsional Alfv\'en waves are generated within a magnetic flux tube embedded in the solar atmosphere. 
The net energy flux carried by these twisters is about $0.2\times 10^4$ W m$^{-2}$. 
This is almost sufficient for the required coronal heating of the quiet-Sun and the solar wind energy losses. 
The vertical flows associated with the magnetic twister carry the mass-flux of its maximum of $6\times 10^{-6}$ kg m$^{-2}$ s$^{-1}$ 
to the inner coronal heights, and thereby they also contribute to the mass supply in the region of nascent solar wind formation. 
The Poynting energy flux carried out by the magnetic twisters is high compared to the energy transported by chromospheric swirls. 
Therefore, we suggest that the magnetic twisters must be considered as an important candidate for the localized coronal heating and the solar wind acceleration. 

The shell solutions seem to be the natural consequence of 
centering the initial pulse on the axis of the tube. 
This ensures that the perturbations have the same symmetry as the tube, and rings/shells are the obvious result. 
For more realistic perturbations, where the symmetry is not exact, multiple vortices with different orientations are likely to be the result 
(see e.g. the 'realistically' driven vortices in Moll et al. 2012, their Figs. 9 and 11). 
For the implications of this on the observed heating see van Ballegooijen et al. (2011).

%
\begin{figure}
\centering{
           \includegraphics[width=8.5cm,angle=0]{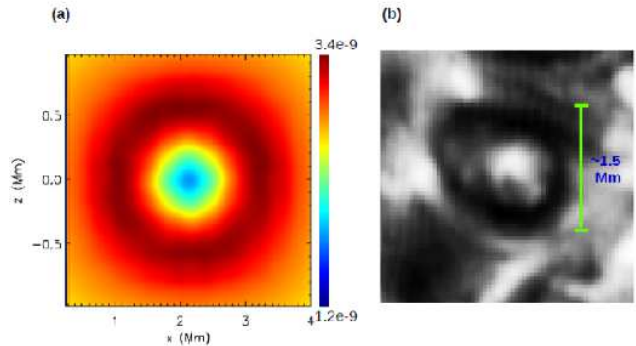}
          }
\caption{\small
         The line-of-sight view of spatially averaged mass density-map in the units of $10^{-9}$ kg m$^{-3}$ 
         (a) at the upper chromospheric height below the transition region is compared to the observational features 
         of recently discovered ring-type chromospheric swirl (Wedemeyer et al. 2013) (b). 
         The comparison shows that recently observed chromospheric swirls may be an integrated appearance of our multi-shell magnetic twisters; 
         note that the spatial resolution of the presented observations is $540$ km per pixel.
        }
\label{fig:compare}
\end{figure}
%
Based on the current observations, the typical morphologies of the swirls are: ring and spiral (Wedemeyer et al. 2013). 
If we fade the resolution of density map as shown in Fig. 8 by 2-3 times, then the multi-shell chromospheric swirls appear 
as a ring type swirl as recently observed in the solar chromosphere (Wedemeyer Bohm \& Rouppe van der Voort, 2009; Wedemeyer-B\"ohm et al. 2012, Wedemeyer et al. 2013). 
The ring-type swirl and its apparent motion are intimately connected with the localized magnetic flux tube, filled-in plasma, 
and their collective dynamics under the influence of the driver. 
The multi-shell magnetic twister will appear like ring-type swirl in a straight flux tube when observed with imager with less spatial resolution. 
This comparison is shown in Fig.~\ref{fig:compare}, where an observed ring-type chromospheric swirl matches in morphology over 
almost similar spatial-scale of modeled faded (density averaged) multi-shell magnetic twister. 
However, our model twister will appear as multi-shell magnetic twister once the spatial resolution is scaled down to less than $70$ km per pixel. 

Therefore, our newly proposed model of multi-shell magnetic twister provides the exact underlying physics of their drivers, morphology, 
and forecast on observations. 
The comparison shows that recently observed chromospheric swirls may be an integrated appearance of our multi-shell magnetic twisters. 
Moreover, the present novel model solves the paradigm of the association of Alfv\'en waves with such twisters. 
It is shown that they are present at multi-shells and collectively carry large amount of energy to heat the corona locally and accelerate the solar wind.

Based on the physical properties of the multi-shell magnetic twisters and the amount of energy and mass carried by them, 
it is suggested that these multi-shell twisters are responsible for the observed heating of 
the solar inner corona and for the formation of nascent solar wind. 
Moreover, it is likely that the existence of the multi-shell magnetic twisters can be verified by observations at higher resolutions.

%
\acknowledgments
We thank the referee for his/her valuable comments and suggestions that allowed us
to significantly improve our paper. 
This work has been supported by NSF under the grant AGS 1246074 (K.M. \& Z.E.M.), 
and by the Alexander von Humboldt Foundation (Z.E.M.). A.K.S. thanks Sven Wedemeyer-B\"ohm 
for providing the observational result of ring-type chromospheric swirl. 
The software used in this work was in part developed by 
the DOE-supported ASC/Alliance Center for Astrophysical Thermonuclear Flashes at the University of Chicago. 
The 2D and 3D visualizations of the simulation variables have been carried out using respectively the IDL (Interactive Data Language) 
and VAPOR (Visualization and Analysis Platform) software packages.
%
%
%
%



\begin{references}
%
%
%
%
\reference{2008ApJS}
Avrett, E.~H., Loeser, R.\ 2008, ApJS, 175, 228
%
%
%
%
\reference{Cranmer}
Cranmer, S.~R.\ 2002, Space Sci. Rev., 101, 229
%
%
%
\reference{De Pontieu}
De Pontieu, B., Erdelyi, R., James, S.~P.\ 2004, Nature, 430,  536
%
\reference{De Pontieu2014}
De Pontieu, B., Rouppe van der Voort, L., McIntosh, S.~W., et al.\ 2014, Science, 346, id. 1255732, 315
%
%
\reference{2000ApJS..131..273F} 
Fryxell, B., Olson, K., Ricker, P., et al.\ 2000, \apjs, 131, 273
%
%
\reference{2013JCoPh.243..269L} 
Lee, D.\ 2013, J. Comp. Phys., 243, 26
%
\reference{2009JCoPh.228..952L} 
Lee, D., Deane, A.~E.\ 2009, J. Comp. Phys., 228, 952 
%
%
%
\reference{MacNeice1999} 
MacNeice, P., Spicer, D.~S., Antiochos, S.\ 1999, 
8th SOHO Workshop: Plasma Dynamics and Diagnostics in the Solar Transition Region and Corona, 446, 457
%
\reference{McIntosh}
McIntosh, S.~W., de Pontieu, B., Carlsson, M., et al.\ 2011, Nature, 475, 477
%
%
\reference{Moll}
Moll, R., Cameron, R.~H., Sch\"ussler, M.\ 2012, \aap 541, 68
%
\reference{2010A&A...518A..37M} 
Murawski, K., Musielak, Z.~E.\ 2010, \aap, 518, A37 
%
\reference{2015A&A} 
Murawski, K., Solov'ev, A., Musielak, Z.~E., Srivastava, A.~K., Kraskiewicz, J.\ 2015, \aap, in press
%
%
%
%
%
\reference{Peter}
Peter, H., Dwivedi, B.~N.\ 2014, Frontiers in Astronomy and Space Sciences, 1-2
%
%
\reference{1988}
Priest, E., Foley, C.~R., Heyvaerts, J., et al.\ 1998, Nature, 393, 545
%
%
%
%
\reference{Shelyag2011}
Shelyag, S., Keys, P., Mathioudakis, M., Keenan, F.\ 2011, \aa, 526, A5
%
\reference{Shelyag2013}
Shelyag, S., Cally, P.~S., Reid, A., Mathioudakis, M.\ 2013, \apjl, 776, L4
%
\reference{Shibata}
Shibata, K., Nakamura, T., Matsumoto, T., et al.\ 2007, Science, 318, 1591
%
\reference{2010A&R}
Solov'ev, A.A.\ 2010, Astronomy Reports, 54, 85
%
\reference{Steiner}
Steiner, O., Franz, M., Bello, G.~N., et al.\ 2010, \apjl, 723, 180
%
\reference{Su}
Su, Y., Wang, T. Veronig, A., et al.\ 2012, \apjl, 756, 41
%
\reference{Tian}
Tian, H., DeLuca, E.~E., Cranmer, S.~R., et al.\ 2014, Science, 346, id. 1255711, 315
%
\reference{2000Toth}
T\'oth, G.\ 2000, J. Comp. Phys., 161, 605
%
%
%
\reference{van}
van Ballegooijen, A.~A., Asgari-Targhi, M., Cranmer, S.~R., DeLuca, E.~E.\ 2011, \apj, 736, 3
%
\reference{2012ApJ} 
Vigeesh, G., Fedun, V., Hasan, S.~S., Erd\'elyi, R.\ 2012, ApJ, 755, 11
%

\reference{Wedeme}
Wedemeyer-Bohm, S., Rouppe van der Voort, L., 2009, A\&A, 507, L9.
%
\reference{Wedemeyer}
Wedemeyer-Böhm, S., Scullion, E., Steiner, O., et al.\ 2012, Nature, 486, 505
%
\reference{Wedemeyer&Steiner}
Wedemeyer, S., Steiner, O.\ 2014, PASJapan, 66, 108
%
\reference{Wedemeyer2013}
Wedemeyer, S., Scullion, E., Steiner, O., et al.\ 2013, J. Phys.: Conf. Ser., 440, 012005
%
\reference{Zhang}
Zhang, J., Liu, Y.\ 2011, \apjl, 741, L7
%
%
\end{references}
\end{document}